%% file: ms.tex
\documentclass{emulateapj}

\usepackage{amsmath}

\shorttitle{Environmental Effects on SF}
\shortauthors{Atlee et al.}

\begin{document}
\title{A Multi-Wavelength Study of Low Redshift Clusters of Galaxies II. 
Environmental Impact on Galaxy Growth}
\author{David W. Atlee\altaffilmark{1} and  Paul Martini}
\affil{Department of Astronomy, The Ohio State University \\
4055 McPherson Laboratory, 140 W. $18^{th}$ Ave., Columbus, OH 43210, USA}
\email{atlee@astronomy.ohio-state.edu}
\altaffiltext{1}{The National Optical Astronomy Observatory, 950 N.\ Cherry Avenue, Tucson AZ 85719}

\begin{abstract}
Galaxy clusters provide powerful laboratories for the study of galaxy
evolution, particularly the origin of correlations of morphology and 
star formation rate (SFR) with density.
We construct visible to MIR spectral energy distributions (SEDs)
of cluster galaxies and use them to measure stellar masses and 
SFRs in eight low redshift clusters, which
we examine as a function of environment.
A partial correlation analysis indicates that SFR depends strongly 
on $R/R_{200}$ ($>99.9\%$ confidence) and is independent of 
projected local density at fixed radius. SFR also shows no residual dependence
on stellar mass.  We therefore conclude that interactions with
the intra-cluster medium drive the evolution of SFRs in 
cluster galaxies.
A merged sample of galaxies from the five most complete clusters shows
$\langle SFR\rangle\propto (R/R_{200})^{1.3\pm0.7}$ for
galaxies with $R/R_{200}\leq 0.4$.  A decline
in the fraction of SFGs toward the cluster center contributes most of this
effect, but
it is accompanied by a reduction in SFRs
among star-forming galaxies (SFGs) near the cluster center.  
The increase in the fraction of SFGs toward larger $R/R_{200}$ and the
isolation of SFGs with reduced SFRs near the cluster center are 
consistent with ram pressure stripping as the mechanism to
truncate star formation in galaxy clusters.  We conclude that stripping
drives the properties of SFGs over the range of radii we examine.
We also find that galaxies near the cluster center are more massive
than galaxies farther out in the cluster at $\sim 3.5\sigma$, which
suggests that cluster galaxies experience dynamical relaxation during the
course of their evolution.
\end{abstract}


\section{Introduction}\label{secIntro}
The current paradigm for the evolution of the
universe and the growth of structure is based largely on observations
of the luminous matter
in the universe, i.e.\ individual galaxies, groups and clusters, and the
cosmic microwave background.
While galaxy formation physics could previously be neglected, the
era of precision cosmology increasingly demands detailed knowledge of
galaxy formation to map observations of luminous
matter onto dark matter halos (e.g.\ \citealt{vDaa11}).
To do so precisely, we
must understand the relationship between galaxy evolution and environment.

Galaxy formation theory dates to the middle of the twentieth century.
Early work explored the physical
processes responsible for star-formation \citep{whip46}, speculated about
the origins of the Milky Way \citep{egge62}, and examined the impact of
environment on galaxy evolution \citep{spit51}.
\citet{oste60} discovered that star-forming galaxies (SFGs) 
are less common in galaxy clusters
than in lower density environments, and this result was subsequently
re-examined with larger samples \citep{gisl78,dres85}.
The dearth of vigorous star formation in galaxy clusters is mirrored by an
under-abundance
of spiral galaxies in these high density regions, known as
the morphology-density relation \citep{dres80,post84,dres97,post05}.

The impact of environment on the frequency and intensity of star-formation
has been studied intensely in galaxy clusters and also
at a variety of other density scales.  These measurements have employed
both visible wavelength colors \citep{koda01,balo04b,bark09,hans09} 
and emission lines
\citep{abra96,balo97,balo00,kauf04,chri05,pogg06,verd08,brag09,vdLi10}
as well as mid-infrared (MIR) luminosities \citep{bai06,sain08,bai09}.  SFGs
are consistently found to be more common and to have higher SFRs
in lower density environments and at higher redshifts
\citep{kauf04,pogg06,pogg08}.  This trend appears to reverse by $z\approx2$,
with star formation more common in clusters than in the field
\citep{tran10,hatc11}.  However, even high-$z$ cluster galaxies form
their stars earlier than coeval field galaxies \citep{rett11}, which
is an expression of the so-called ``downsizing'' phenomenon \citep{cowi96}.

The relationships between SFR, morphology and environment in the local
universe place strong constraints on models for galaxy evolution.
Another important factor is
the presence of an evolutionary trend for galaxies to have higher SFRs
at higher redshifts.  This was originally reported as an excess of blue
cluster members at
$z\approx0.4$ compared to $z=0$ \citep{oeml74,butc78,butc84}, and is
commonly known at the Butcher-Oemler Effect.  
This trend is now understood to track the simultaneous increase in the
fraction of star-forming galaxies (SFGs) and in the SFRs of individual
SFGs.  An analogous trend
has been examined in the MIR
\citep{sain08,hain09,tran10,hatc11}, which is sensitive to
dust-enshrouded star formation.

The observed trends in star formation with environment and the variation
of these trends with redshift are
usually attributed to changes in the sizes of cold gas reservoirs.
Several mechanisms have been proposed to reduce galaxies' cold gas
supplies and transform them from SFGs to passive galaxies.
These mechanisms include ram-pressure stripping of cold gas
(RPS; 
\citealt{gunn72,abad99,quil00,roed05,roed06a,roed07,jach07}), 
gas starvation
\citep{lars80,balo00,bekk02,kawa08,mcca08,book10},
galaxy harassment \citep{moor96,moor98,lake98}, and interactions
with the cluster tidal potential \citep{merr83,merr84,nata98}.
Gas starvation operates throughout clusters, and it
converts galaxies from star forming to passive
on a gas exhaustion timescale, which, for normal spiral galaxies,
is $\sim 2.5$~Gyr \citep{bigi11}.  This time is similar to the cluster
crossing time of 2.4~Gyr, which is the timescale appropriate for dynamical
processes like galaxy harassment.  These timescales
contrast sharply with the timescale
appropriate for RPS, which truncates star formation on a gas stripping
timescale, which is of order $10^{5}$~yr.  The efficiency of RPS also scales
with ICM density, so it operates much more strongly near cluster centers than
either starvation or harassment.
These differences mean that how rapidly star formation declines in cluster
galaxies relative to the field
constrains the mechanism primarily responsible for the removal of
cold gas from cluster galaxies.

The variation of SFR with environment can probe the relative importance
of different environmental processes, but the conclusions drawn from
apparently similar observations sometimes conflict.
For example, \citet{mora07} identified passive spirals in
a sample of $z\approx0.5$ clusters and determined that spiral galaxies
rapidly turn passive when they enter the cluster environment 
and then evolve into S0 galaxies.
\citet{bai09} argue that the similarity
of the $24\mu m$ luminosity functions observed in galaxy clusters
and in the field suggests that the transition from star-formation to
quiescence
must be rapid, which implies that ram pressure stripping (RPS) is the
dominant mechanism.  \citet{verd08} and \citet{vdLi10},
by contrast, find a significant
trend of increasing SFR with radius to at least 
$2 R_{200}$ from cluster
centers.  Because the trend of SFR with radius appears to extend to the 
virial shock \citep{whit91}, 
\citet{vdLi10} conclude that preprocessing at the group scale is
important.  
\citet{pate09} find a similar trend for increasing average SFR with
decreasing local density down to group-scale densities
($\Sigma_{gal}\approx1.0\ {\rm Mpc^{-2}}$).

Evidence for pre-processing in groups is important because
RPS is inefficient in low density gas, so preprocessing 
\citep{zabl98,fuji04} is likely driven by processes 
like gas starvation that operate in less dense environments.
While preprocessing appears to be important in some groups and clusters, 
\citet{berr09} found that
very few cluster galaxies have previously resided in groups, so the impact
of preprocessing on a typical cluster galaxy must be limited.

In Atlee et al.\ (2011; henceforth Paper I), we developed a multi-wavelength
method to identify AGNs based on either their X-ray luminosities or the
shapes of their visible-MIR spectral energy distributions (SEDs).  
In this paper, we will employ this method to correct for AGNs and measure
star formation in cluster galaxies.  We use the results to study
the relationship between star formation and the cluster environment.
In particular, we consider the constraints placed on the important 
environmental processes that operate in clusters by the distribution of 
star formation among cluster members.

The paper is organized as follows: In \S\ref{secMethods}, we review
our observations, which are discussed in more detail in Paper I.
In \S\ref{secPartialMath} we review the mathematical formalism associated
with partial correlation analysis.
In \S\ref{secComplete} we derive completeness corrections for the observed
cluster members.  We discuss the derivation of total
infrared (TIR) luminosity functions (LFs) in \S\ref{secLF}, and
in \S\ref{secResults} we detail the results of our measurements.
Finally, we examine the implications of these results for the environmental
dependence of galaxy evolution in \S\ref{secDiscuss}, 
and we summarize these conclusions
in \S\ref{secConclusion}.
Throughout this paper we adopt the WMAP 5-year cosmology---a $\Lambda$-CDM
universe with $\Omega_{m}=0.26$,
$\Omega_{\Lambda}=0.74$ and $h=0.72$ \citep{dunk09}.

\section{Observations and Member Description}\label{secMethods}
Paper I provides details of our photometry and spectroscopy.  It also
develops methods to reliably identify low-luminosity AGNs
and to measure galaxy properties like stellar mass and SFRs
for identified cluster members, including AGN hosts.
We briefly summarize the salient points below.

\subsection{Observations}\label{secObs}
We identified cluster member galaxies using redshifts determined by 
\citet{mart07}.  We supplemented these with redshifts from the literature,
which we obtained from the 
NASA Extragalactic Database (NED)\footnote{http://nedwww.ipac.caltech.edu/}.
(These redshifts 
come from a variety of sources with unknown
selection functions and success rates.  See \S\ref{secComplete}.)

We have visible-wavelength images from the
du Pont telescope at Las Campanas observatory, MIR images from
the IRAC and MIPS instruments on the {\it Spitzer Space Telescope}, and
X-ray images extracted from the {\it Chandra} archive.
We measure visible, MIR and X-ray fluxes 
in redshift-dependent photometric apertures that approximate
a fixed metric size.  The aperture fluxes are then corrected to total
fluxes at constant color with the $R$-band Kron-like magnitude
from SExtractor \citep{bert96}.
Our photometry spans
the peak of the stellar continuum, so we have robust photometric
redshifts, which can identify catastrophic errors among the 
spectroscopic redshifts.
We found 12 such catastrophic errors, one of which is an AGN (Paper I).  
We exclude these objects from our analysis.

\subsection{Cluster Member Description and AGN Identification}\label{secSeds}
In Paper I, we constructed spectral energy distributions (SEDs) from the 
photometry described in \S\ref{secObs}, and we fit models to these fluxes with
SED template codes from \citet{asse10}.
These model SEDs are used to derive photometric redshifts and $K$-corrections.
We also employed the models to measure MIR color
corrections, and we used the results to determine rest-frame luminosities.

We employ rest-frame, visible wavelength 
colors to determine the mass to light ratio of each galaxy in the 
sample \citep{bell01}, and we combine these with the measured luminosities
to infer $M_{*}$.  To measure SFRs, we employ both $8\mu m$
and $24\mu m$ luminosities \citep{zhu08}.  When SFRs can be
measured in both bands, we take the geometric mean of the two measurements.
We found in Paper I that
SFRs determined independently from $8\mu m$ and $24\mu m$ luminosities 
show a scatter of $\sim0.2~{\rm dex}$ with respect to one another.  This 
scatter reflects the systematic uncertainty in SFR measurements determined from
either band separately, and the implied uncertainty is comparable
to the systematic uncertainties in the measured stellar masses (
0.3 dex, Paper I).

Before we measure stellar masses and SFRs of cluster members, we
identify and correct for AGNs.  We employ two independenty methods to identify 
AGNs: the shapes of the model SEDs (IR AGNs) and the X-ray luminosities 
measured with {\it Chandra} (X-ray AGNs).  The hosts of IR AGNs
are corrected for the presence of the AGN before we calculate $M_{*}$
and SFR.  X-ray AGNs without visible signatures in their SEDs 
do not contribute significantly
to the measured visible and MIR fluxes, so we do not correct those
objects.  In Paper I, we found that the IR and X-ray AGN samples are largely
disjoint.  This implies that X-ray only AGN selection
can overlook a large fraction ($\sim 35\%$) of AGNs.
We explore the potential consequences of this
bias in \S\ref{secSubstr} and \S\ref{secBO}.

\section{Partial Correlation Analysis}\label{secPartialMath}
When confronted with a system of mutually correlated observables, it can be
difficult to establish which variables drive the correlations.
Partial correlation analysis measures the relationship between 
two variables with all other parameters held fixed and can identify which
variable(s) control the observed correlations.
Partial correlation analysis has been applied in the
past to develop a fundamental plane of black hole activity \citep{merl03} and
to probe the dependence of SFR on both stellar mass and environment 
simultaneously \citep{chri05}.  We will use the simplest formulation of
partial correlation analysis, which relies only on direct measurements and
does not account for upper limits.

The simplest case
is a system of only three variables, $x_{i}$.  This is
called the first-order partial correlation problem.
The correlation coefficient for $x_{1}$ and $x_{2}$ at fixed $x_{3}$ can
be expressed as
\begin{equation}\label{eqFirstPartial}
r_{12.3}=\frac{\rho_{12}-\rho_{13}\rho_{23}}{ \sqrt{(1-\rho^{2}_{13})(1-\rho^{2}_{23})} }
\end{equation}
where $\rho_{ij}$ is the standard two-variable correlation coefficient 
(e.g.\ the Pearson or Spearman coefficients) between $x_{i}$
 and $x_{j}$ \citep{wall03}.  Higher
order problems describe systems with more variables.
For a system of $N$ variables, the $(N-2)^{th}$ order
partial correlation coefficient 
$r_{ij.1...N \backslash \{ij\}}$ of variables $x_{i}$ and $x_{j}$ 
can be written,
\begin{equation}\label{eqPartialCovar}
r_{ij.1...N \backslash \{ij\}}=\frac{-C_{i,j}}{ \sqrt{C_{i,i} C_{j,j}} }
\end{equation}
where $C_{i,j}=(-1)^{i+j}M_{i,j}$ \citep{kend77}.
$M_{i,j}$ is a reduced determinant of the
correlation matrix $R$, where $R_{i,j}=\rho_{ij}$, and $\rho_{ij}$
is the two-variable correlation coefficient of $x_{i}$ and $x_{j}$.  
The determinant $M_{i,j}$
can be interpreted as the total correlation among the variables of the system
in the absence of $i$ and $j$.  It
is calculated from $R$ with the $i^{th}$ 
row and $j^{th}$ column eliminated \citep{kend77}.

Given a partial correlation coefficient from Eq.\ \ref{eqPartialCovar},
we would like to know its significance.  This
can be evaluated from $\sigma_{ij.1...N\backslash \{ij\}}$,
\begin{equation}\label{eqSigmaPartial}
\sigma_{ij.1...N\backslash \{ij\}}=\frac{1-r_{ij.1...N \backslash \{ij\}}}{ \sqrt{m-N} }
\end{equation}
where $r_{ij.1...N\backslash \{ij\}}$ is the 
partial correlation coefficient given by Eq.\ 
\ref{eqPartialCovar}, $N$ is the number of variables in the system, and
$m$ is the number of objects in the sample.  The statistical significance
of $r_{ij.1...N\backslash \{ij\}}$ is determined from
the Student's t-distribution with dispersion
$\sigma_{ij.1...N\backslash \{ij\}}$ \citep{wall03}.

Partial correlation analyses can take both parametric and non-parametric
forms.  These are analogous to the more commonly applied two-variable
correlation analyses.  
Equation \ref{eqPartialCovar} can be applied to any of the correlation 
coefficients in
common use.  However, Eq.\ \ref{eqSigmaPartial} is defined for the parametric
Pearson's correlation coefficient, so it is appropriate only for that
estimator or the closely related, non-parametric Spearman coefficient.
We want a non-parametric approach, so we rely on Spearman 
correlation coefficients in our analysis.

\section{Completeness Corrections}\label{secComplete}
We wish to examine the distributions of the stellar masses and 
SFRs described in \S\ref{secSeds} to
probe the impact of the cluster environment on galaxy growth.  However, to
do this we must first correct for selection effects.  The spectroscopic 
selection function that defines our sample is unknown, because
many of the sources that contribute to the redshifts in the literature
do not define their
target selection functions or rates of success.
Furthermore, the MIR observations
do not uniformly cover the cluster fields.  Therefore, we empirically
determine both our spectroscopic and MIR selection functions
to correct for these effects.

\subsection{Spectroscopic Completeness}\label{secSpecComp}
We examine only spectroscopically-confirmed cluster members.  Many of the
redshifts we use come from \citet{mart07}, 
which we supplemented with redshifts from other sources in the literature.
This results in a complex selection function that is poorly known
{\it a priori}.  However, this completeness function is required to correct
the properties of observed cluster galaxies to the intrinsic distribution
for all cluster members.  We take an empirical approach to determine 
spectroscopic completeness and correct the measured cluster members to the
total cluster galaxy population.

For each cluster, we bin galaxies identified in the 
photometric source catalog
by $V-R$ color, $R$-band magnitude and
$R/R_{200}$.  We find significant
variations in the fraction of galaxies with spectra ($f_{spec}$)
as a function of $R/R_{200}$\ and $m_{R}$,
but the variation with color is at most minor.  A partial
correlation analysis of $f_{spec}$ as a function of color, magnitude 
and position shows no significant partial
correlation with $V-R$ at 95\% confidence in any cluster, while $f_{spec}$
correlates with both $m_{R}$ and $R/R_{200}$ at $>99.9\%$\ confidence.  We 
therefore collapse the measurement along the color axis and determine
the fraction of galaxies with spectroscopy as a function of $R$-magnitude and
position only.  This results in better measurements due to the larger
number of galaxies per bin.

The $f_{spec}$ measured above is one way to express the
spectroscopic completeness of galaxies in a given magnitude-radius bin.  
However, what we really want is
an expression for the spectroscopic completeness, $C_{spec}$,
of cluster members,
\begin{eqnarray}\label{eqSpecComp}
C_{spec}(\vec{x}) & = & \frac{N_{Cl,spec}(\vec{x})}{N_{Cl}(\vec{x})} \\
              & = & \frac{N_{spec}(\vec{x})}{N_{tot}(\vec{x})} \times
                    \frac{N_{Cl,spec}(\vec{x})}{N_{spec}(\vec{x})} \times 
                    \frac{N_{tot}(\vec{x})}{N_{Cl}(\vec{x})}
\end{eqnarray}
where $\vec{x}$ is the position of a given bin in magnitude-radius space,
$N_{Cl,spec}$ is the number of galaxies with spectra that are cluster members,
$N_{Cl}$ is the number of true cluster members, 
$N_{spec}$ is the number of galaxies
with spectra in the cluster field, and $N_{tot}$ is the number of galaxies
in the input catalog.  All of these quantities except $N_{Cl}$ can be measured 
directly from the input catalogs.  We would need to infer $N_{Cl}$ using
some additional piece of information, so we prefer to rely on $f_{spec}$
rather than $C_{spec}$ if possible.

\begin{figure}
\epsscale{1.0}
\plotone{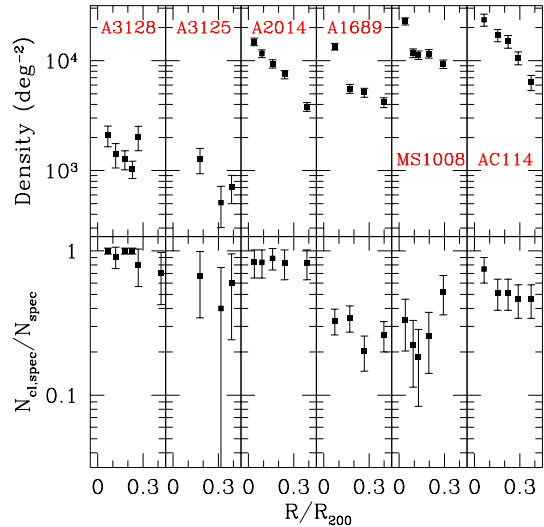}
\caption{Comparison of galaxy density ({\it upper panel}) and 
spectroscopic membership fraction ({\it lower panel})
as a function of radius for the clusters in our sample with enough 
confirmed members to make a meaningful measurement.  If $f_{spec}$ were a
good proxy for $f_{Cl,spec}$, the upper and lower panels would have similar
slopes.  None of the clusters in the sample exhibit this behavior.
\label{figMemRad}}
\end{figure}

If the redshifts reported in the literature were not pre-selected for cluster
membership
or if the redshift failure rate was high, $f_{spec}(\vec{x})$ would
be a good proxy for $C_{spec}(\vec{x})$, and the approach in
Eq.\ \ref{eqSpecComp} would be unnecessary.
If this were the case, the fraction of galaxies
with spectra that are cluster members ($f_{mem}$) should drop with $R/R_{200}$
as the fraction of field galaxies increases.  Figure \ref{figMemRad} shows
that $f_{mem}$ does not always
trace the decline in the density of cluster galaxies.  This implies that
$f_{spec}$ is not a good tracer of $C_{spec}$, and the more sophisticated
approach of Eq.\ \ref{eqSpecComp} is required.  

Before we can employ
Eq.\ \ref{eqSpecComp}, we need to know the number of cluster galaxies in
each bin.  To do this, we estimate the number of field galaxies in the bin
with the $R$-band magnitude-number density relation reported by 
\citet{kumm01}.  We
subtract the field galaxies from the total number of galaxies in the bin 
to estimate the
number of cluster galaxies.  

This approach introduces two types of uncertainty.
The first is simple Poisson counting uncertainty due to the
small number of field galaxies, typically a few to 10, in each bin.  The
second is cosmic variance.  \citet{elli87} reports a $B$-band magnitude-number
relation that includes measurements from a number of other authors.  The
different surveys use fields of different sizes, so the scatter of their
results about the best-fit relation provide a measure of the cosmic variance,
which contributes of order 10\% uncertainty on the
number of field galaxies in a typical bin.  The number of field galaxies
in a given bin depends on magnitude and cluster mass, but it generally
ranges from 1-10 galaxies.  At faint magnitudes, the number of field
galaxies is generally comparable to the number of cluster galaxies, and
Poisson fluctuations in the number of field galaxies dominate the 
uncertainties in the completeness measurements.  The spectroscopic
completeness measurements and associated uncertainties for each cluster 
are summarized in Table 
\ref{tabComplete}.

\begin{figure}
\epsscale{1.0}
\plotone{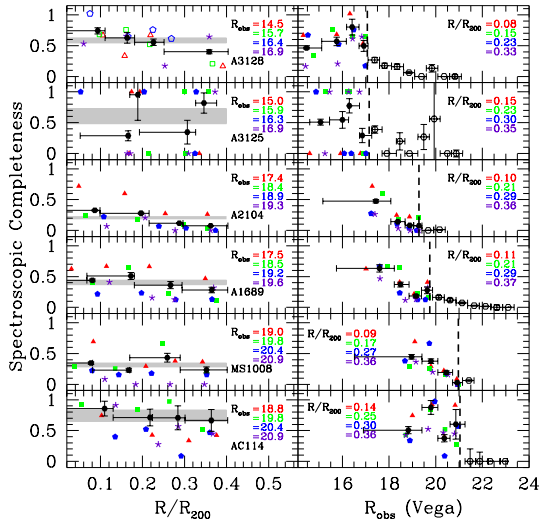}
\caption{Fraction of cluster galaxies with spectra
as a function of projected distance from the cluster center ({\it left}) 
and $R$-band magnitude ({\it right}).  The measurements in each column
have been separated by the indicated variable (colored points).
The black points show the completeness averaged over all of the
colored bins, and the grey bands show the $1\sigma$ confidence intervals
on the total completeness in each cluster.  
The dashed, vertical lines on the righthand column
indicate the magnitude corresponding to $M_{R}\approx-20$~mag.
The average measurements use only galaxies with $M_{R}\leq-20$~mag and
$R/R_{200}<0.4$ ({\it filled, black points}).
\label{figSpecComp}}
\end{figure}

Figure \ref{figSpecComp} shows the spectroscopic completeness ($C_{spec}$)
for 6 of the 8 galaxy clusters in our
sample.  The remaining 2 clusters (A644 and A2163) have too few confirmed
members
to make a reliable measurement.  The dashed, vertical lines on the right 
column of 
Figure \ref{figSpecComp} indicate the
observed magnitude that corresponds to $M_{R}=-20$ for the average 
K-correction in each cluster.  The follow-up spectroscopy of
X-ray sources conducted by \citet{mart06} is only complete to this
luminosity limit.  
Clearly, completeness becomes quite poor for $M_{R}>-20$ in all clusters,
so we restrict our sample to galaxies with $M_{R}<-20$.

We also considered completeness as a function of luminosity and stellar mass
instead of $m_{R}$.  However, these quantities have higher uncertainties than
observed magnitudes, especially for galaxies without spectroscopic redshifts
to fix their distances.  Therefore, we measure completness as a function
of $m_{R}$ and $R/R_{200}$.

\subsection{Mid-Infrared Completeness}\label{secMirComp}
The depth of the MIR images varies
as a function of position across the clusters.  This is a result of
the {\it Spitzer} mosaicking schemes, which were chosen to provide
good coverage of
the known X-ray point sources in the cluster.  These mosaic schemes lead to
variations in the number of overlapping images, and therefore to variations
in sensitivity, across the cluster fields.

In addition to these sensitivity variations, the {\it Spitzer}
footprint features
some non-overlapping coverage by the different IRAC bands.  This results
from the IRAC mapping strategy, which simultaneously images two
adjacent fields in different bands.
The pointings chosen by the observer then determine
the degree of overlap between the IRAC channels.  For
a galaxy to enter the final sample, it must include detections in at least
5 bands to ensure that the fit results for that galaxy are well constrained.
This means that a faint galaxy in a region of a
cluster with overlapping $3.6\mu m$\ and $4.5\mu m$\ images, for example,
might be more likely to appear in the final sample than an identical galaxy
in a part of the cluster with only $4.5\mu m$ coverage.

\begin{figure}
\epsscale{1.0}
\plotone{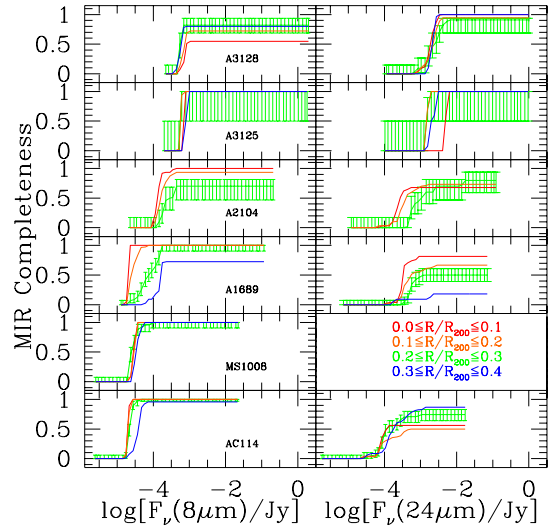}
\caption{MIR completeness as a function of flux for the $8\mu m$\ 
({\it left}) and the $24\mu m$\ ({\it right}) 
{\it Spitzer} bands.  In each column, the sample has
been separated into 4 radial bins.
Fluxes have not been color-corrected and are given in
the observer frame.  Uncertainties are shown for a single radial
bin to indicate typical values.
MS 1008.1-1224 was not observed with MIPS. 
Completeness measurements are derived as described 
in \S\ref{secMirComp}.
\label{figMirComp}}
\end{figure}

To construct ensemble statistics for whole clusters, we require sensitivity
corrections that account for variable depth across the cluster fields and 
for the different footprints in the {\it Spitzer}
bands.  We again take an empirical
approach to completeness correction.
We measure the MIR flux uncertainties at the locations of all
confirmed cluster members from the {\it Spitzer}
uncertainty mosaics.
At each position, we combine the two \citet{asse10} star-forming templates
with arbitrary flux normalizations 1000 times to produce galaxies with
$10^{-2}<SFR/1\ M_{\odot}\ yr^{-1}<10^{2}$.  From these artificial galaxy SEDs,
we construct model fluxes and determine whether the galaxy represented 
by each SED would have been
detected at $3\sigma$ based on the flux uncertainty at each
position.  We bin the results by flux
and by $R/R_{200}$ to estimate completeness separately at $8\mu m$ and
$24\mu m$.  Figure 
\ref{figMirComp} shows the results of this measurement for the 6 clusters
in Figure \ref{figSpecComp}.
IRAC and MIPS completenesses clearly depend on both flux and $R/R_{200}$.

The uncertainties in MIR completeness result from the
incomplete spectroscopic sampling of the galaxies in a given bin.
We implicitly assume that the identified cluster members in each bin 
are representative of
the behavior of the unidentified members.  This assumption means that the
precision of the completeness correction in a given bin is fixed by the
number of identified cluster members in that bin.  The
completeness we measure, $C_{MIR}$, is our best estimate of the
``true'' MIR completeness $C_{MIR}^{true}$ associated with a hypothetical,
spectroscopically complete sample of galaxies.
In a bin with $N$ cluster members, the expected number
of detections is simply $C_{MIR}^{true} N$.  However, the actual number of 
detections will have some range around $C_{MIR}^{true} N$,
which leads to an uncertainty in the
inversion of $C_{MIR}$ to a completeness correction.  
This uncertainty is set by the expected variation
in the number of galaxies, which is best described by binomial statistics.
This allows calculation of asymmetric error bars on $C_{MIR}$ and
accounts naturally for upper and lower limits.  Typical uncertainties
returned by this procedure are $\sim 20\%$.
The full set of MIR completeness measurements and the associated 
uncertainties are 
summarized in Table \ref{tabMirComplete}.

\subsection{Merged Cluster Sample}\label{secStack}
We have defined the MIR completeness measurements in Figure \ref{figMirComp}
so they apply only to galaxies with spectroscopic redshifts.  The
two corrections, applied serially, give total completeness corrections.
The total correction $X_{G}$ for a galaxy $G$ is,
\begin{equation}\label{eqCompCorr}
X_{G} = \frac{1}{C_{spec}(R_{G}/R_{200},m_{R,G})}\times \frac{1}{C_{MIR}(R_{G}/R_{200},f_{\nu,G})}
\end{equation}
where $C_{spec}$
is the spectroscopic completeness (Figure \ref{figSpecComp}) and
$C_{MIR}$ is the MIR completeness (Figure \ref{figMirComp}).  The
completeness corrections described by Eq.\ \ref{eqCompCorr} can be
applied to individual galaxies to extrapolate from the measured
galaxy samples to the full cluster population.
In cases where multiple corrections can be derived for a single object,
we combine these
corrections in the same way the data are combined.  For example,
the completeness correction for a galaxy with SFR measurements
from both $8\mu m$ and $24\mu m$ fluxes is given by
$X = \bigl[ X_{8\mu m}X_{24\mu m}\bigr] ^{1/2}$ because
$SFR = \bigl[ SFR_{8\mu m}SFR_{24\mu m}\bigr] ^{1/2}$
(\S\ref{secSeds}).

To examine the dependence of star formation and black hole growth on
environment, we need to construct a merged cluster galaxy sample.  We identify
5 clusters (A3128, A2104, A1689, MS1008 and AC114) with the best completeness
estimates and combine their members.  The relatively small number of
galaxies in A3125 results
in highly irregular behavior of the completeness functions.  As a result,
any corrections we applied would depend critically on the binning scheme.
We therefore exclude it from the main cluster sample.

The clusters in the main sample can be stacked to yield
better statistics.
To construct the stacked cluster, we weight individual galaxies by their 
completeness corrections ($X_{G}$).  The correction is a combination
of the spectroscopic and photometric completeness corrections
from \S\ref{secSpecComp}
and \S\ref{secMirComp}, as given by Eqs.\ \ref{eqSpecComp}-\ref{eqCompCorr}. 

\section{Luminosity Functions}\label{secLF}
Luminosity functions (LFs) provide an important diagnostic for the
difference between cluster galaxies and field populations, because
LFs are sensitive to the entire cluster
population rather than only the average.
For example, \citet{bai09} employed the total infrared (TIR) LF
to infer that RPS controls the evolution of SFRs in cluster galaxies.
In this section, we discuss the derivation of total infrared (TIR) luminosities
and our method to construct luminosity functions.  We discuss the
results in \S\ref{secTirLF}.

\subsection{Total Infrared Luminosity}\label{secBolCorr}
Our MIR observations cover a relatively narrow wavelength range from $3.6\mu m$
to $24\mu m$.  To compare our results with previous studies,
we need to infer $L_{TIR}$ from the observed $L_{8\mu m}$ and $L_{24\mu m}$, 
so we must apply
bolometric corrections (BCs).
To estimate $L_{TIR}$ from the {\it Spitzer} luminosities, we
employ the \citet{dale02} SED template library, which
includes a wide variety of SEDs.  These SEDs differ
from one another according to the parameter $\alpha$, which describes the
intensity of the radiation field on a typical dust grain.

Before we calculate BCs for IR AGNs, we first subtract the AGN contribution
(Paper I).
We then fit each \citet{dale02} template to the rest-frame
$5.8$, $8.0$ and $24\mu m$ fluxes
and use the template that best fits the data to measure our fiducial BCs.
In the frequent cases where luminosities in one or more
of these bands are unavailable, we estimate the missing luminosities from
model SEDs \citep{asse10}.  When this is necessary, we assign 
uncertainties to the model fluxes from the uncertainties on the model SED.
We only calculate $L_{TIR}$ for galaxies with detections in at least
one of the $8\mu m$ and $24\mu m$
bands.

In galaxies that have measurements of both $L_{8\mu m}$ and $L_{24\mu m}$,
we calculate $L_{TIR}$ separately for each band and take the geometric mean
of the results.  This follows our treatment of SFRs in Paper I.
In other cases, we simply use the BC appropriate for
the band where we have a detection.  Typical BCs are $\sim 6$ for $L_{8\mu m}$
and $\sim 8$ for $L_{24\mu m}$.
We also construct 68\% confidence intervals
for each BC based on the $\Delta \chi^{2}=1$ interval for each galaxy.  
These uncertainties are asymmetric, and they
add in quadrature to the uncertainties on $L_{8}$
and $L_{24}$ to give the total uncertainty on $L_{TIR}$.

\subsection{Luminosity Function Construction}\label{secLfConstruct}
We construct luminosity functions from galaxies in the main cluster
sample whose luminosities---$L_{TIR}=L^{+\sigma_{u}}_{-\sigma_{l}}$---we
determine as described in \S\ref{secBolCorr}.
If we account for the uncertainties on $L_{TIR}$, we can reduce our 
sensitivity to Poisson fluctuations in the number of luminous galaxies.  
We distribute the galaxy weights described in \S\ref{secComplete} over 
luminosity bins according to the probability that the true luminosity of 
a galaxy with best-estimate
$L=L_{TIR}$ lies in a given bin.  Due to the uncertainty on the LF prior, 
this technique increases the statistical uncertainty on the total weight
in each bin by $\sim10\%$.
In exchange, we reduce the much larger uncertainty introduced by 
stochasticity in the number luminous galaxies.

To distribute galaxy weights over luminosity bins, we employ an asymmetric 
probability density function (PDF)
that considers $\sigma_{l}$ and $\sigma_{u}$ separately.
We integrate the PDF across each luminosity bin to determine the weight 
in each bin, which we add to construct the total LF.
The PDF we employ is piecewise smooth, and it approaches 
the Normal distribution when $\sigma_{u}\approx\sigma_{l}$.
It is described in more detail in Appendix
\ref{appAsymGaus}. 

In addition to the PDF, we require a prior 
on the shape of the LF to correct for Eddington-like bias due to the steepness
of the LF above $L_{*}$.  We adopt a Schechter function fit
to the Coma cluster LF from \citet{bai06} as
the baseline prior.  We then correct the Coma LF to the redshift
of each individual cluster according to the evolution of the field galaxy
LF \citep{lefl05}.  We add the uncertainty on the prior to the 
statistical uncertainty on the LF in each luminosity bin.
The prior has a
strong impact on the bright-end shape of the LF because there are few
cluster galaxies to constrain the LF in this regime.
The results are discussed in \S\ref{secTirLF}.

\section{Results}\label{secResults}
We apply the weights derived from the completeness corrections
described in \S\ref{secComplete} to the main cluster sample, which is
a subset of the cluster galaxies in
Table \ref{tabGalaxies}.  These corrections allow us to examine the
average environmental dependence of $M_{*}$ (\S\ref{secMassRad})
and SFR (\S\ref{secSfrRad}-\ref{secSubstr}), and 
the redshift dependence of star formation
(\S\ref{secBO}).  Before conducting these analyses, we perform
a partial correlation analysis to determine which observed properties
of galaxies in clusters most strongly correlate with star formation
(\S\ref{secPartial}).  The results inform the rest of our work.

\subsection{Partial Correlation Analysis}\label{secPartial}
\begin{figure}
\epsscale{1.0}
\plotone{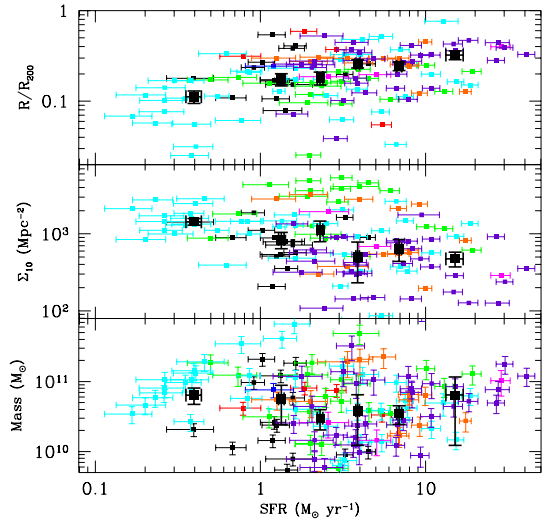}
\caption{Correlations of star formation with position in the cluster
(top row), projected local density (middle row), and stellar
mass (bottom row).  Galaxies with no measurable
star formation are neglected.  Colors denote the different clusters in
the sample: A3128 ({\it black}), A3125 ({\it red}), A644 ({\it blue}), 
A2104 ({\it green}), 
A1689 ({\it cyan}), A2163 ({\it magenta}), MS1008 ({\it orange}), AC114 
({\it violet}).  Large black points show the median values of the galaxy
sample after it has been binned by SFR.  SFR shows strong correlations
with both $R/R_{200}$
and $\Sigma_{10}$ ($r_{S}=+0.35$ and $r_{S}=-0.34$, respectively), but
no correlation with $M_{*}$.
Partial correlation coefficients derived from these data are listed in
Table \ref{tabSfCorr}.
\label{figEnvCorr}}
\end{figure}

The cluster environment can significantly alter the evolution of cluster
member galaxies, as described in \S\ref{secIntro}.  However, when we attempt
to distinguish between the physical properties that might cause these
effects, we confront a system of mutually-correlated observables.
For example, SFR depends on both
projected local galaxy density to the $10^{th}$ nearest neighbor
($\Sigma_{10}$; 
\citealt{oste60,oeml74,dres80,kauf04}) and
position within the cluster ($R/R_{200}$; 
\citealt{koda01,balo04b,chri05,blan07b,hans09,vdLi10}).
Figure \ref{figEnvCorr} demonstrates the correlations between
SFR, position, projected galaxy density and $M_{*}$ among SFGs. 
It is not immediately clear which of these is the most fundamental.

While the causal connection between morphology and the local density of
galaxies is well established
(e.g.\ \citealt{dres80,dres99,post05}), \citet{mora07} find strong
evidence that the morphologies and star-formation rates of massive
spiral galaxies in clusters evolve separately.  This implies that a factor
other than local density may control star formation in cluster galaxies.
Because $M_{*}$, $R/R_{200}$ and projected local density ($\Sigma_{10}$)
are all mutually correlated,
it is not easy to determine which variable(s) drive the environmental
dependence of star formation.  Therefore, we use a partial correlation 
analysis to disentangle these dependencies.
The mathematical formalism for partial correlation analysis is described in
\S\ref{secPartialMath}.  
We do not consider completeness corrections for this analysis, so
we include galaxies from all 8 clusters.

We consider only objects with measurements of
all parameters under consideration and ignore galaxies with upper limits.
This differs from the similar analysis conducted by \citet{chri05}, who also
considered upper limits.
As a result, our results are more sensitive than
\citet{chri05} to systematic effects like variations in sensitivity
within or between clusters.  Because of this, we do not rely directly on the 
strength of any partial correlations, but only on the presence or
absence of such correlations.  For variables with significant 
partial correlations, we perform stacking analyses, which can account
for incompleteness.  (See \S\ref{secMassRad}-\S\ref{secBO}.)

We perform a partial correlation study on a system of five variables:
SFR, $M_{*}$, $R/R_{200}$, the \citet{dres88} substructure
parameter ($\delta$), and projected local density of cluster members
($\Sigma_{10}$).
The partial correlation coefficients returned by the analysis are listed
in Table \ref{tabSfCorr}.

Table \ref{tabSfCorr} shows that SFR depends strongly on 
$R/R_{200}$ ($r_{S,partial}=+0.34$), but it shows
no significant dependence of SFR on $M_{*}$ once the influence of $R/R_{200}$
has been factored out ($r_{S,partial}=+0.09$).  This conflicts with
earlier results, which generally find either that SFR depends only on $M_{*}$
\citep{grut11,rett11} or that SFR depends on both $M_{*}$
and environment \citep{chri05}.  One reason for this discrepancy is
that Table \ref{tabSfCorr} does not include non-SFGs.  The fraction of
non-SFGs is higher among more massive galaxies, which introduces a
dependence of $\langle SFR\rangle$ on $M_{*}$.  Another possible factor
is the difference in stellar masses we examine.  The samples of \citet{grut11}
and \citet{rett11} extend down only to $\sim 3\times10^{10}M_{\odot}$, while
our sample extends to approximately $10^{10} M_{\odot}$.  If the cluster 
environment affects lower mass galaxies more strongly, we should be
more sensitive to this effect.  Because the efficiency of RPS should depend
on $M_{*}$, our expanded mass range improves our ability to distinguish 
between RPS and gas starvation.

Interestingly, SFR also shows no residual dependence on
$\Sigma_{10}$\ ($r_{S,partial}=+0.02$), which
suggests that the SFRs of cluster members are driven
by the local conditions of the ICM rather than by
interactions with nearby galaxies.  
We emphasize that this conclusion applies to SFR only, and we do not
consider the dependence of morphology on $\Sigma_{10}$ or $R/R_{200}$.
This result is also consistent with
results from previous authors \citep{pogg99,mora06,mora07} who found that
the processes that alter SFR and morphology are likely to be physically
distinct.
The determination that SFR is more closely related to $R/R_{200}$ than
to $\Sigma_{10}$ distinguishes
our results from those of \citet{chri05}, who do not discriminate between
different environmental tracers.  As a result, they are agnostic about the
process(es) that drive the $\langle SFR\rangle$--radius relation.
Our analysis also relies on SFGs alone, which distinguishes it from the
work of \citet{chri05}, who included upper limits for galaxies
with no measurable star formation.  This accounts for the
lack of a strong anti-correlation between $M_{*}$ and SFR, which is driven
by a decline in the fraction of SFGs at higher $M_{*}$ rather than a
reduction in the SFRs of individual SFGs.

SFR also shows no relationship to local substructure, as measured by $\delta$,
at fixed $R/R_{200}$.
Indeed, even a two-variable correlation
test returns no correlation between $\delta$\ and SFR 
($r_{S}=-0.03$).
This conflicts with \citet{chri05}, who reported
a strong correlation of SFR with local substructure.  However, $\delta$
requires a robust spectroscopic sample from which to measure local
velocity dispersions.  As a result, the substructure measurements for
some of our clusters with less complete spectroscopy are probably
unreliable.  We repeat the test with A3128 and A3125, which have the most
complete spectroscopy and 
are the only two clusters with significant substructure.  The results are
indistinguishable from the full cluster sample.

\subsection{Mass--Radius Relation}\label{secMassRad}
In \S\ref{secPartial}, we reported a strong correlation of SFR with
$R/R_{200}$ but no residual dependence of SFR on $M_{*}$ or $M_{*}$ on radius.
This might indicate that $M_{*}$ is independent of environment, as
\citet{vdLi10} found.  However, it might
also mean that the SFR--$R/R_{200}$ correlation is strong enough to eclipse any
more subtle correlations that might appear among a sample composed entirely
of SFGs.
The galaxy sample examined in \S\ref{secPartial}
includes only a few hundred galaxies, and the sample preferentially excludes
the most massive galaxies, which tend not to show active star formation.
As a result, \S\ref{secPartial} might show no correlation between $M_{*}$
and $R/R_{200}$, even if the full cluster galaxy sample includes one.
\citet{chri05} found a strong partial correlation
of mass with $R/R_{200}$.  This correlation would be difficult to
produce if BCGs alone produce
a false correlation of $M_{*}$ with $R/R_{200}$,
as \citet{vdLi10} claim, because
normal cluster galaxies are much more numerous than BCGs.

\begin{figure}
\epsscale{1.0}
\plotone{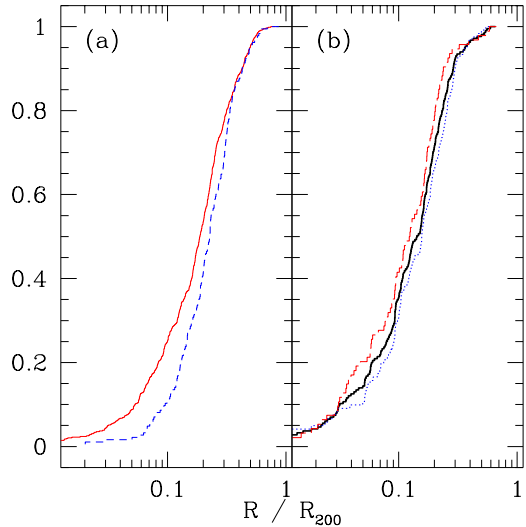}
\caption{Radial distributions of all cluster members scaled to
$R_{200}$.  Panel (a) compares the radial distributions of
low-mass ({\it blue dashed})
and high-mass ({\it red solid}) galaxies, divided into two equally-sized
subsamples at $M_{cut}=3.9\times10^{10}\ M_{\odot}$.  The two distributions
differ at 99.9\% confidence after we exclude BCGs as defined by
\citet{vdLi07}.  Panel (b) compares the radial
distributions of galaxies
with ({\it blue dotted}) and without ({\it red dashed}) an
$8\mu m$ flux excess to the distribution of all galaxies with
$8\mu m$ detections ({\it heavy black}).
The distribution of galaxies with no measurable
excess shows a marginal difference compared to the distribution of 
all cluster members (95\% confidence).  
\label{figRadialDist}}
\end{figure}

\begin{figure}
\epsscale{1.0}
\plotone{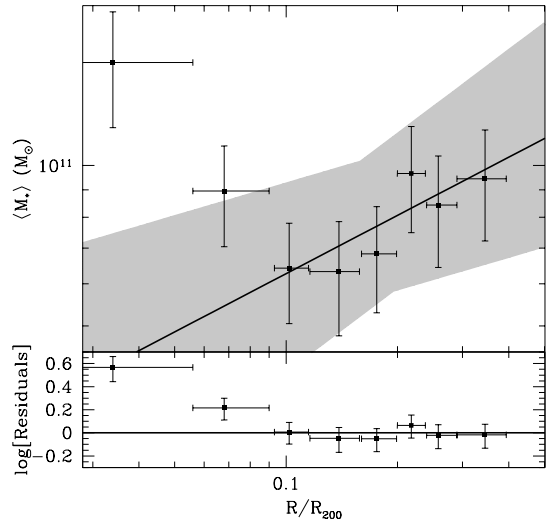}
\caption{Average stellar mass for galaxies in the stacked cluster sample as a
function of radius.  Brightest cluster galaxies are excluded from
the fit because of their unusually large stellar masses and SFRs
compared to other galaxies near the centers of clusters.
The heavy line indicates the best-fit power-law to the 6 outer bins.  The
two innermost radial bins were excluded from the fit based on their large
excesses.  This resulted in a reduction in total $\chi^{2}$ from 
$9.1$ to $1.0$.  The best
fit yields $M_{*}\propto\bigl[R/R_{200}\bigr]^{0.4\pm0.2}$,
and the shaded region indicates the 68\% confidence interval 
to the fit.  The residuals
are shown in the lower panel.  
\label{figMassRad}}
\end{figure}

To test whether our data support the presence of a radial trend in
$M_{*}$, we look for direct variations of $M_{*}$
with $R/R_{200}$ without regard to correlations with other variables.
We first divide the galaxy sample into two samples with equal numbers of
galaxies, and we apply a K-S test to check for a difference between their
radial distributions.  For this
analysis, we include members of all 8 clusters, and we exclude BCGs as defined
by \citet{vdLi07} from the sample.  The results
are shown in Figure \ref{figRadialDist}a.
The K-S test returns a probability $<0.1\%$ that the high- and low-mass
samples have the same radial distributions, so massive galaxies are 
preferentially found closer to the centers of their parent clusters,
even in the absence of BCGs.  We weight members of the
main cluster sample by their completeness to determine
the average mass as a function of radius.  
The average mass in a given bin is,
\begin{equation}\label{eqAvgMass}
\langle M_{*}\rangle = \frac{\Sigma^{N}_{i=0} \bigl[w_{i} M_{*,i}\bigr]}{\Sigma^{N}_{i=0} \bigl[ w_{i}\bigr]}
\end{equation}
where $N$ is the number of galaxies in the bin with $M_{R}<-20$.
The $w_{i}$ are the weights derived from Eq.\ \ref{eqSpecComp}.  
Figure \ref{figMassRad}
shows the resulting average masses as a function of radius.  

The innermost radial bin Figure \ref{figMassRad} shows a strong excess
compared to the other bins, and the second bin hints at an excess.
We fit a power law to the six outer radial bins in
Figure \ref{figMassRad} (solid line) 
to measure the strength of the mass excess in
these bins and to determine why our results differ from \citet{vdLi10}.
The slight increase of $\langle M_{*}\rangle$ with $R/R_{200}$ 
beyond $R\approx 0.1R_{200}$ is consistent
with the tendency of massive galaxies to be accreted more recently than less
massive galaxies and for recently accreted galaxies to lie further out
in the cluster \citep{deLu11c}.  
Alternatively, galaxies presently near the center of the
cluster spend more time near the cluster center,
on average, than galaxies farther away.  Therefore, they are subject to
stronger tidal forces from the cluster potential and will 
lose more of their mass \citep{merr83,merr84,nata98}.  
The large uncertainties on the
fit preclude any attempt to distinguish between these scenarios.

Figure \ref{figMassRad} shows that the average masses of galaxies
near the cluster center show an excess compared to their counterparts
further out in the cluster.  The innermost radial bin in Figure
\ref{figMassRad} differs from the best fit model by $2.8\sigma$,
and the mass excess in the second radial bin is significant at $1.8\sigma$.
This indicates that the cluster core ($R\lesssim 0.05R_{200}$)
tends to host more massive galaxies
than the outer regions, even if we neglect BCGs.  Mass segregation among
cluster galaxies can be introduced as the cluster relaxes to virial
equilibrium.  The dynamical relaxation timescale in the inner mass bin
for a cluster at $z=0.15$ with $\sigma=1200~km~s^{-1}$ is approximately 
300 Myr, so the core of a typical cluster in our sample should be dynamically
relaxed.  For the same ``typical'' cluster, the crossing time for the sphere
defined by $R\leq 0.4R_{200}$ is approximately 900 Myr.
This crossing time implies a dynamical relaxation time of 7 Gyr, which is
longer than the age of a cluster at $z=0.15$ that ``assembled'' at $z=1$
(5.8 Gyr), so the sphere with $R\leq 0.4R_{200}$ has not yet had time to relax.

Finally, Figure \ref{figMassRad} suggests a reason for our disagreement
with \citet{vdLi10}.  The signal comes primarily inside $0.05 R_{200}$,
which corresponds to $\sim1\farcm{5}$ at the median redshift of the 
\citet{vdLi10} sample ($z\approx0.8$).  Due to SDSS fiber collisions, only
a few galaxies inside $0.05 R_{200}$ will have redshifts in each cluster.
This shifts the median of the innermost radial bin in the \citet{vdLi10}
sample to $\sim 0.08 R_{200}$.  This is comparable to the second radial bin
in Figure \ref{figMassRad}.  If this was our innermost bin, we would not 
find any dependence of $M_{*}$ on $R/R_{200}$, so the disagreement
between our results and \citet{vdLi10} likely result from fiber
collisions in SDSS.

\subsection{Environmental Dependence of SFR}\label{secSfrRad}
In Paper I, we examined the $R/R_{200}$ distributions of AGNs and found no
significant difference between the positions of AGNs and normal cluster 
members.  The lack of radial dependence among AGNs could
due to the small sample size, it could indicate a weak
dependence of the amount of cold gas on $R/R_{200}$,
or it might mean that AGN fueling is poorly correlated with the total cold gas 
reservoir of its host galaxy.  To test these hypotheses, we define a sample
of galaxies with $8\mu m$ flux excesses as those galaxies whose measured
$8\mu m$ flux exceeds the flux expected from a passively evolving
galaxy matched in $M_{K}$ at more than $2\sigma$.
Figure \ref{figRadialDist}b compares the
radial distributions of galaxies with and without an $8\mu m$ excess.
These objects include both SFGs and AGNs.  We again excluded BCGs 
from these samples.  The radial distribution of 
galaxies with $8\mu m$ excesses is indistinguishable from
the merged sample, but galaxies without an excess are located
closer to the centers of their host clusters than the average cluster galaxy
at 95\% confidence.

The dependence of dust emission on $R/R_{200}$ shown in 
Figure \ref{figRadialDist}b is consistent
with the established dependence of SFR on position within
galaxy clusters \citep{koda01,balo04b,chri05,hans09,vdLi10} and with our
results in \S\ref{secPartial}.  One way to test the origin of this
effect, and by extension the
SFR--density and SFR--radius relations,
is to measure the average SFR as a function of radius.  We weight individual
SFGs by their total completeness (Eq.\ \ref{eqCompCorr}) and bin them
in radius to determine $\langle SFR\rangle$:
\begin{equation}\label{eqAvgSfr}
\langle SFR\rangle = \frac{\Sigma^{N_{SF}}_{i=0} \bigl[w_{SF,i} SFR\bigr]}{\Sigma^{N_{gal}}_{j=0} \bigl[ w_{j}\bigr]}
\end{equation}
where $w_{SF,i}$ and $w_{j}$ are the weights for SFGs and all galaxies, 
respectively.  For this calculation, we define SFGs as all galaxies
with $SFR\geq 3~M_{\odot}~yr^{-1}$.  This guarantees that we are not
subject to biases due to variable sensitivities across the cluster fields.

\begin{figure}
\epsscale{1.0}
\plotone{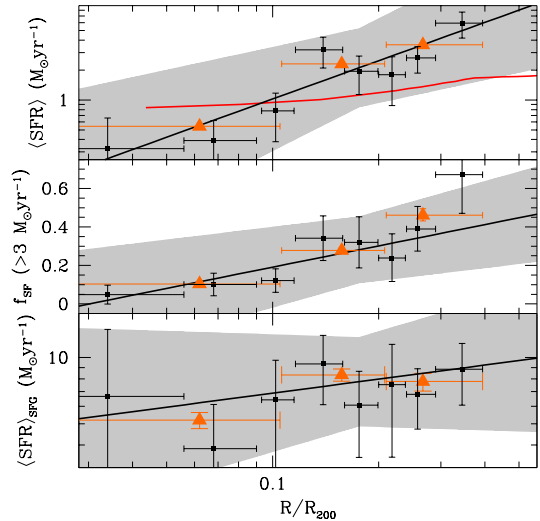}
\caption{Average star formation in the stacked cluster sample 
as a function of position.  Each panel shows two different binning
schemes: fine (\textit{black squares}) and coarse (\textit{orange triangles})
for the same galaxy samples.
The {\it top panel} shows average SFR among all galaxies.  The   
solid line indicates the best-fit power law to the
data $\bigl(SFR\propto\bigl[R/R_{200}\bigr]^{1.3\pm0.7}\bigr)$,
and the \textit{red line} shows the gas starvation model of \citet{book10},
normalized to match the observed SFRs.
The shaded region indicates the $1\sigma$ confidence interval 
to the fit.  The {\it middle panel} shows the fraction of SFGs
($SFR>3\ M_{\odot}\ yr^{-1}$) as a function of position,
with the best fit $\bigl(f_{SF}\propto [R/R_{200}]^{0.4\pm0.2}\bigr)$
shown by the line and the $1\sigma$ confidence interval shown by the shaded 
region.
The {\it bottom panel} shows the averaged SFR among SFGs 
($\langle SFR\rangle_{SFG}$) 
versus $R/R_{200}$.  
Galaxies with $R/R_{200}\lesssim 0.1$ have lower 
$\langle SFR\rangle_{SFG}$ than galaxies outside $0.1 R_{200}$
at $>99.9\%$ confidence.
\label{figSfRad}}
\end{figure}

We fit a power law to $\langle SFR\rangle$ as a function of $R/R_{200}$
and find,
\begin{equation}\label{eqSfRad}
\log_{10}\bigl[SFR\bigr]=(1.3\pm0.7)\log_{10}\bigl[R/R_{200}\bigr] + (1.3\pm0.6)
\end{equation}
where $SFR$ is the average in each radial bin.  The fit yields
$\chi^{2}_{\nu}=1.2$ and is shown by the solid line on the upper panel
with the $1\sigma$ uncertainty given by the grey region.  The red line
shows the gradient predicted by the gas starvation model of \citet{book10},
normalized to minimize the $\chi^{2}$ between the model and the observed 
SFRs.  The model predicts
the average SFR for all non-BCG cluster members at $z=0$, so the higher
redshift of the cluster galaxies in Figure \ref{figSfRad} might explain the
shift in $\langle SFR\rangle$ between the observations and the model.
However, even after we adjust the normalization of the model, it remains
a poor fit to the data ($\chi^{2}_{\nu}=3.2$), but the 
re-normalized model is marginally consistent with the best-fit power law within
the large statistical uncertainty on the fit.

Despite the significant uncertainty in the power law fit to the SFR--radius
relation, Figure \ref{figSfRad}
clearly demonstrates higher $\langle SFR\rangle$ toward the outer regions 
of the stacked cluster sample.  This mirrors the trend found via partial
correlation analysis in 
\S\ref{secPartial}, which is significant at $>$99.9\% confidence.  A
similar trend appears in $\langle sSFR\rangle$ as a function of radius, 
which yields $\langle sSFR\rangle \propto (R/R_{200})^{1.0\pm0.6}$.
To interpret Eq.\ \ref{eqSfRad} in detail, we need a 
model that better agrees with the observations than the
\citet{book10} model and that accounts for projection effects, 
the distribution of orbits 
followed by cluster members, and the effect of different environmental 
processes.

The RPS
scenario makes at least one clear, qualitative prediction that we can use
to evaluate its impact without a detailed model.
Because RPS operates quickly compared to the cluster crossing 
time, the radial variation in $\langle SFR\rangle$ should be caused by
variations in the fraction of SFGs ($f_{SF}$), and there should be little
change in the SFRs of individual galaxies. 
The middle panel of Figure \ref{figSfRad} shows that $f_{SF}$ declines strongly
near the cluster center, which is consistent with RPS.  
We also see lower
$\langle SFR\rangle$ among SFGs with $R\lesssim 0.1~R_{200}$ compared
to SFGs with $R>0.1R_{200}$ (bottom panel).
This difference is formally significant at $>99.9\%$ confidence,
so SFGs with $R\lesssim 0.1R_{200}$ experience a clear reduction
in their SFRs as they transition to passive evolution.  However, a fit
to $\langle SFR\rangle_{SFG}$ versus $R/R_{200}$ shows no significant trend, so
the reduction in SFR at $R\lesssim 0.1R_{200}$ may be a sharp transition
rather than a gradual decline.  The time to cross this region is approximately
100 Myr, which is consistent with the hypothesis that the apparent break is
due to the influence of RPS.  The onset of this break, however, occurs
much closer to the cluster center than is usually expected 
($\sim 0.5 R_{200}$; \citealt{treu03}).

The small region over which RPS has the strongest effect might explain the
absence of the correlation between $M_{*}$ and SFR that is expected under
the RPS scenario.  The most massive galaxies, which are best able to retain
their gas, are preferentially found at $R\lesssim 0.1R_{200}$.  These
galaxies therefore experience stronger ram pressure than typical, less
massive cluster members, and their ability to retain their gas is cancelled
by the increased stripping they experience.

\subsection{TIR Luminosity Function}\label{secTirLF}
Another probe of the impact of environment on star formation is
the TIR luminosity function (LF).  The TIR LF is sensitive to the frequency of
star formation in clusters and the rapidity with which it is quenched;
this provides a strong empirical constraint on the types of processes that
mediate the interaction between individual galaxies and the cluster
environment.  For example, \citet{bai09} found similar
shapes ($\alpha$\ and $L^{*}$)
of the TIR LFs in the galaxy clusters that they measure compared to
the field galaxy TIR LF.  They argue that this similarity requires
truncation of star formation on short timescales compared to
the lifetime of star formation in individual galaxies.
Such rapid transitions are inconsistent with
processes like gas starvation and galaxy harassment.

\begin{figure}
\epsscale{1.0}
\plotone{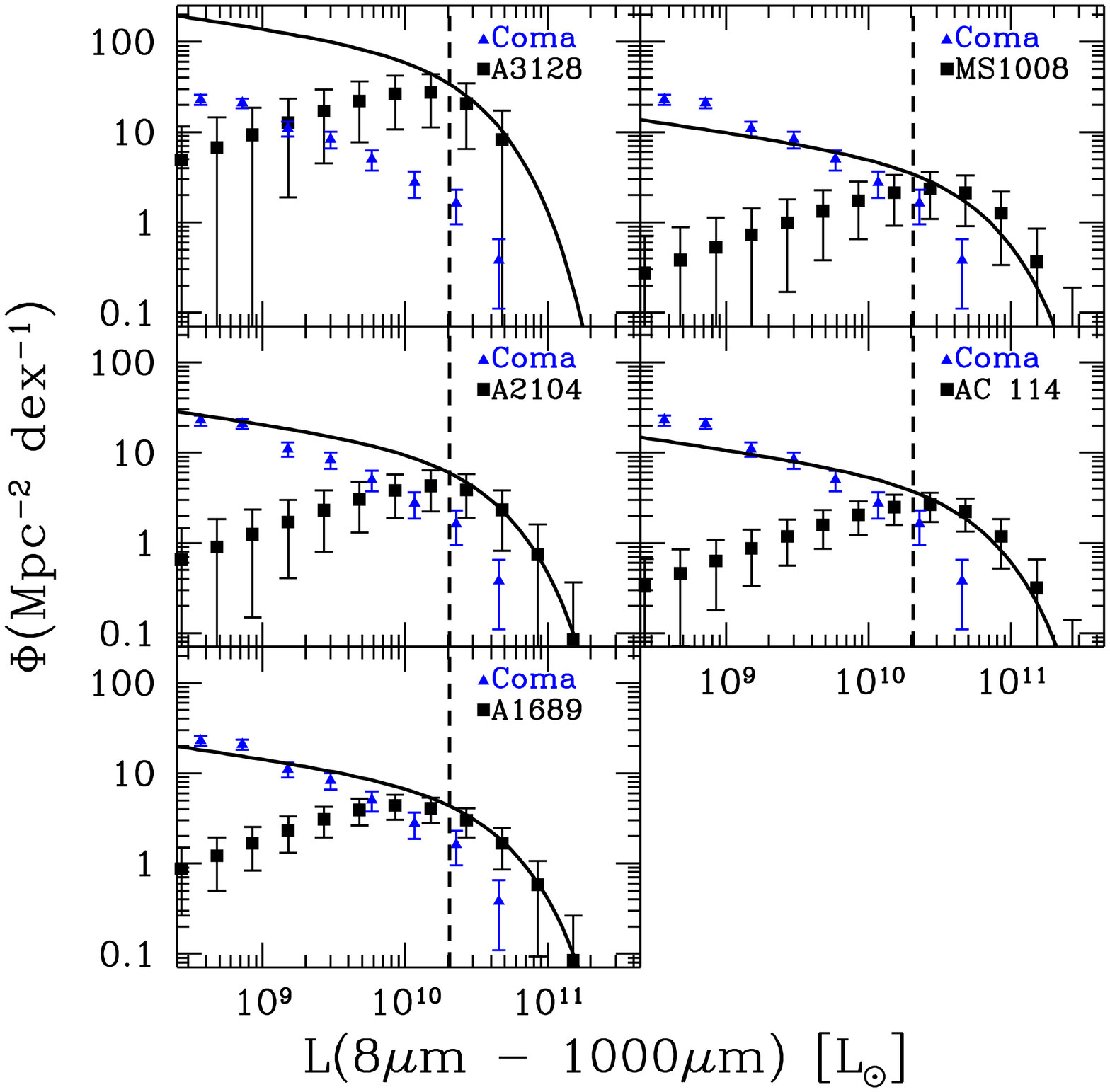}
\caption{Total infrared (TIR)
luminosity functions for each of the 5 clusters in the main sample.
Input galaxies are restricted to $M_{R}<-20$ and $R<0.4R_{200}$.
The Coma cluster LF ({\it blue triangles};
\citealt{bai09}) is shown for reference.  The \textit{solid black lines}
on each panel show the field galaxy TIR LF of \citet{pere05} shifted to the 
redshift of the cluster and normalized to match the observed LF above
our nominal completeness limit of $2.1\times 10^{10} L_{\odot}$ (vertical, 
dashed line), which is the approximate $L_{TIR}$ expected for a normal, spiral
galaxy with $M_{R}=20$~mag.
This indicates the expected distribution of SFRs among
field galaxies that enter the cluster.  The \textit{dashed lines}
mark the completeness limit imposed by the requirement $M_{R}<-20$.
Some variation between individual clusters is apparent at the highest 
luminosities.
\label{figClusterLFs}}
\end{figure}

To evaluate the conclusion that RPS dominates the evolution of
star formation in galaxy cluster members, 
we will examine the TIR LFs of the clusters in 
our main cluster sample.  We construct the LF as described in
\S\ref{secLfConstruct}, and the results appear in Figure \ref{figClusterLFs}.
The main cluster sample contains only 5 clusters, which prevents
construction of subsamples that have different masses
and similar redshifts.  Therefore, we cannot reliably identify effects that are
strongly dependent on cluster mass.

The dashed vertical line
in Figure \ref{figClusterLFs} marks the expected $L_{TIR}$
of a galaxy with the \citet{asse10} spiral SED and $M_{R}=-20$.  
This marks the approximate TIR completeness limit imposed by the requirement 
that $M_{R}\leq -20$.  We call this limit $L_{TIR}^{thresh}$.
This limit is representative only, and Figure 
\ref{figClusterLFs} includes
many cluster members that have $M_{R}\leq -20$ and $L_{TIR}<L_{TIR}^{thresh}$.
This is expected because cluster galaxies
have lower $\langle sSFR\rangle$ than the field galaxies used to
construct the \citet{asse10} templates.  In fact, 65\% of galaxies with 
$M_{R}<-20$~mag and 
measurable ($>3\sigma$) MIR emission are less luminous than
$L_{TIR}^{thresh}$.  This means that $L_{TIR}^{thresh}$ is robust, and the
true limit
is lower than the nominal value established from the 
spiral galaxy template.  To predict the true $L_{TIR}^{thresh}$,
we would need a model for the truncation of star formation in clusters, which
is exactly what we want to measure.  To be conservative, we restrict our
fits to use only bins more luminous than $L_{TIR}^{thresh}$.
Above this limit, we can be confident that the weights given by 
Eq.\ \ref{eqCompCorr} will
correct to the full galaxy population.

\begin{figure}
\epsscale{1.0}
\plotone{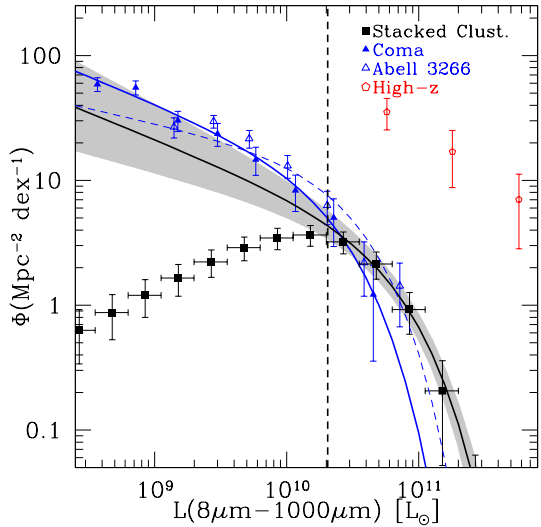}
\caption{TIR LF of the stacked cluster sample ({\it filled squares})
compared to the LFs
derived by \citet{bai09} for Coma ({\it filled triangles}) 
and Abell 3266 ({\it open triangles}).
The solid lines show the best fit LFs for the stacked sample ({\it black})
and for Coma ({\it blue}), and the shaded region shows the 68\%
confidence interval around the best fit to the stacked cluster LF.  The
{\it dashed line} shows the field galaxy LF of \citet{pere05}, and we have
employed the prescription of \citet{lefl05} to correct the LF to the
median redshift of the combined cluster sample.  The field galaxy LF is
normalized to best match the observed LF above the nominal completeness
limit.
The average LF determined by \citet{bai09} from two high-$z$ clusters
(MS 1054-03 and RX J0152, {\it red pentagons})
is shown for comparison.
\label{figStackedLF}}
\end{figure}

Like \citet{bai09}, we find that the individual clusters in Figure
\ref{figClusterLFs} have luminosity functions that closely resemble the
field galaxy LF at their respective redshifts.
This agreement occurs despite the disagreement 
between the field galaxy LF and the combed cluster sample 
(Figure \ref{figStackedLF}).  While it is possible that the field galaxy
LF at the median redshift of the cluster provides a poor estimate of 
$L^{*}_{TIR}$ among galaxies that fall into clusters, any effects of
preprocessing in the large-scale structure around the cluster should appear
in Figure \ref{figClusterLFs}.
A better explanation
appears to be that the improved statistics in the combined cluster sample
illuminate a discrepancy that is not visible in the individual clusters due
to larger observational uncertainties.

At least some of the variation between the IR LFs observed in different
clusters may be caused by systematic uncertainties in the completeness
corrections.  We can only apply completeness corrections in
regions of the clusters where we have both spectra and MIR photometry, so
azimuthal asymmetry may be important.
The LF of the stacked cluster sample averages over several selection regions,
so it is less subject to this uncertainty.

We constructed the best-fit Schechter functions to both the Coma cluster and
the stacked cluster shown in Figure \ref{figStackedLF}.  
The Schechter function has the form,
\begin{equation}\label{eqSchechter}
\Phi(L)=\frac{\Phi^{*}}{L_{*}} \biggl[\frac{L}{L_{*}}\biggr]^{\alpha} e^{-L/L_{*}}
\end{equation}
where $\Phi(L)$ gives the projected surface density of sources
at TIR luminosity L, and $\alpha$
and $L_{*}$ are the usual Schechter function parameters.
We fixed $\alpha=-1.41$ in the fit to the cluster LF, which is the 
best-fit value for the Coma LF \citep{bai06}.
\citet{lefl05} suggest that 
the faint end of the LF cannot evolve much with redshift for
$z\lesssim 1$, so the faint end of the LF in the Coma cluster
is likely to provide a good estimate of $\alpha$ in all galaxy clusters.
The best fit to the stacked main sample has
$L_{*}=(6.6\pm1.1)\times 10^{10} L_{\odot}$.

If clusters rapidly shut off star formation in galaxies that fall in from the
field, as \citet{bai09} conclude, then only galaxies that have recently become
cluster members will have measurable star formation, and the TIR LF of a 
cluster should have $L^{*}$ and $\alpha$ similar to the field galaxy
LF at the same redshift.  Therefore, 
we want to compare $L_{*}$ to the field galaxy LF at the 
median redshift of the combined galaxy sample, $z_{med}=0.24$.  
\citet{lefl05} found that the field galaxy LF
evolves as $L_{*}\propto(1+z)^{n}$, where $n=3.2^{+0.7}_{-0.6}$.
\citet{pere05} studied the $12\mu m$ LFs of field galaxies from $z=0$ to
$z=3$ and found that the field galaxy LF at $z=0.1$ has 
$L_{12\mu m}^{*}=4.1\pm1.3 \times 10^{9}L_{\odot}$ and $\alpha=1.23\pm0.07$.  
We use the prescription of
\citet{take05} to convert their $L^{*}$ to a TIR luminosity, which yields
$L_{TIR}^{*}=2.3\times 10^{10}L_{\odot}$ at $z=0.1$.
We determine the field galaxy LF at the median redshift of the stacked
cluster sample ($z_{med}=0.211$) with the results of \citet{lefl05}
and fit the normalization of the LF to the observed cluster galaxy LF.
The result is shown as the blue, dashed line in Figure \ref{figStackedLF}.
The quality of the fit ($\chi^{2}=4.7$) is considerably poorer than the
fit shown by the heavy, black line ($\chi^{2}=0.5$), which uses 
$\alpha$ from the Coma cluster and fits for $L_{TIR}^{*}$ and $\phi^{*}$.
While the absolute $\chi^{2}$ values cannot be used to evaluate the quality
of the fits due to the presence of correlated errors in adjacent bins,
the Coma-based LF improves the quality of the fit by $\Delta\chi^{2}=4.2$ with
only 1 additional degree of freedom.

\citet{bai09} found that the luminous ends of the TIR LFs of the
Coma cluster and A3266 have similar shapes and that the $L_{TIR}^{*}$ for
these clusters are indistinguishable from the field galaxy LF.
The similarity between the stacked cluster LF and the redshifted field
galaxy LF for $L_{TIR}>4\times10^{10}L_{\odot}$ is consistent with the
conclusions of \citet{bai09}.
They argue that the similar LF shapes in clusters and in the field
suggests that gas starvation is not a plausible 
mechanism to end star formation among cluster member galaxies.  
Because gas starvation operates slowly ($\sim{\rm Gyr}$ timescales), they
conclude, it
should produce many galaxies in the transition phase between SFGs and
passive evolution.  They found no such transition population.
However, the combined cluster LF shown in
Figure \ref{figStackedLF} displays a $4\sigma$ deficit
of galaxies with moderate SFRs ($SFR\approx 5~M_{\odot}~yr^{-1}$) 
compared to the expectation from the
field LF.  This largely drives our conclusion that an unmodified field 
galaxy LF does not provide a good
description of the SFG population in the merged cluster
sample.  The disagreement between the field and cluster galaxy LFs 
could indicate the 
presence of a transition population.
It is possible that the discrepancy results instead from some
selection effect not accounted for in our completeness estimates.  The
most obvious culprit for such an effect
is some residual dependence of spectroscopic
member identification on color that we have not been able to identify from
the data.

\begin{figure}
\epsscale{1.0}
\plotone{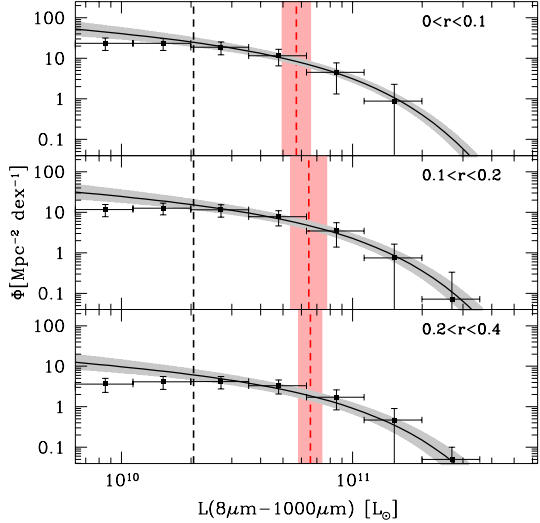}
\caption{TIR LF divided into radial bins with equal numbers of
galaxies.  Each panel is labeled with the range of radii that contribute
to the LF, where $r=R/R_{200}$.
The solid lines show the best-fit Schechter function to the
LF in each radial bin, and the shaded regions show the uncertainties on the
fit.  The fits are constructed from the bin more luminous than the
completeness limit ({\it dashed line}), which is set by the expected 
$L_{TIR}$ for the \citet{asse10} spiral galaxy template at $M_{R}=-20$~mag.  
The {\it red, dashed lines} on each
panel show the best-fit $L_{TIR}^{*}$, and its $1\sigma$ confidence interval
is given by the shaded region.
The three LFs hint at an increase in $L_{TIR}^{*}$ analogous to the
increase in $\langle SFR\rangle_{SFG}$ seen in Figure
\ref{figSfRad}, but this increase is not statistically significant. 
\label{figRadialLF}}
\end{figure}

As a test for radial gradients in the population of transition 
SFGs, we binned the galaxies in the main
cluster sample into three radial bins with equal numbers of
galaxies.  The TIR LFs for the radial subsamples are shown in Figure
\ref{figRadialLF}.  The $L_{TIR}^{*}$ increases slightly from the innermost
to outermost radial bins, but this increase is not statistically significant.
This marginal decrease in $L^{*}_{TIR}$ in the innermost radial bin is 
qualitatively similar to the results of \citet{bai09}, who also examined the
radial dependence of the TIR LF with a very similar binning scheme.
However, the fractional change in $L_{TIR}^{*}$ in the Coma cluster is
much larger than observed in Figure \ref{figRadialLF}.
The increase in $\langle SFR\rangle_{SFG}$
seen in Figure \ref{figSfRad} in the same radial bins
does not appear to affect $L^{*}_{TIR}$.
This indicates
that the change in $\langle SFR\rangle_{SFG}$ is driven by the
frequency of galaxies with $SFR\approx 3~M_{\odot}~yr^{-1}$.

\subsection{Substructure and Preprocessing}\label{secSubstr}
Von der Linden et al.\ (2010) found a trend toward increased 
$\langle SFR\rangle$
at larger $R/R_{200}$ that extended out to at least $2R_{200}$.
They concluded that preprocessing in groups contributes significantly 
to the SFR--density
relation.  Our observations do not extend past $R=0.4R_{200}$, so it
is impossible to measure preprocessing directly, but
A3128 shows significant substructure, so we can compare it to the smooth
clusters in the sample to probe how the presence of substructure 
influences SFRs in clusters.
This allows us to indirectly test the impact
of group scale environments on SFGs, because coherent substructures
in clusters should correspond to recently-accreted groups.

Before we can compare the integrated SFRs in different clusters, we must
first correct for the different numbers of galaxies in each cluster.
The sSFR naturally accounts for this variation,
and it is therefore a better parameter to compare integrated SFRs between
clusters.  We employ a method analogous to Eq.\ \ref{eqAvgSfr}
to calculate the $\langle sSFR\rangle$ and compare A3128,
which shows significant substructure, to the other clusters in the
main sample.  We find 
$\langle sSFR\rangle=9.0^{+1.1}_{-1.2}\times10^{-12}~yr^{-1}$ and
$\langle sSFR\rangle=3.1^{+0.2}_{-0.2}\times10^{-11}~yr^{-1}$ in A3128 and in
clusters without substructure, respectively.
If we correct $\langle sSFR\rangle$ of A3128 to the mean redshift of the
other clusters ($z=0.241$), we find 
$\langle sSFR\rangle=1.5^{+0.4}_{-0.4}\times10^{-11}~yr^{-1}$.  This
is still lower than the average of the clusters without substructure at
$>99.9\%$ significance.

The difference between A3128 and the other clusters might be a result of
the structure in A3128, or A3128 might simply have an unusually low
$\langle sSFR\rangle$ for its redshift.  In the latter case, the observed
difference would be a result of cosmic variance.  We compared A3128 with the
4 individual clusters without substructure, and we found that A3128 has
higher redshift-corrected $\langle sSFR\rangle$ than MS1008, which
is approximately 50\% more massive than A3128.  However, the typical dispersion
in $f_{SF}$ among nearby clusters with $\sigma\gtrsim 800~km~s^{-1}$
is $\sim 0.1$~dex \citet{pogg06}, so to explain the observed deficit of
$\langle sSFR\rangle$ in A3128 as cosmic variance would require a
$\sim 3\sigma$ excursion.  While cosmic variance provides a marginally
plausible explanation for the deficit of $\langle sSFR\rangle$ in A3128,
the presence of substructure appears to be the more likely cause.  
If the observed difference arose from groups that have recently fallen into
the cluster, the excess 
sSFR in clusters without substructure would imply that
the ``average'' group member is likely to have experienced preprocessing.
This result may be absent from the partial correlation results
(Table \ref{tabSfCorr}) because only $\sim 10\%$ of cluster members have ever 
been part of a large group \citep{berr09}, and only former group members that
have been accreted more recently than the $e$-folding timescale for star 
formation will show evidence of preprocessing.

The conclusions drawn from the measured SFRs of cluster members and from
the integrated cluster properties
depend strongly on the methods used to identify and correct for AGN.
In Paper I, we noted that the IR and X-ray AGN selection techniques identify
quite different samples.  If we relied only on X-ray based AGN selection,
as some authors do, the MIR luminosity contributed by unidentified AGN would
lead to an overestimate in the integrated SFR of the cluster.  For example,
in A1689 we would overestimate the total SFR by
20\%.  Applied to all clusters simultaneously, this alternative method
of AGN correction results in an inferred 
$\langle sSFR\rangle=7.2^{+1.5}_{-1.3}\times10^{-11}\ yr^{-1}$ among the
clusters without measurable substructure but no
measurable change in A3128.
In this example, uncorrected AGN contamination would
dominate the observed difference in $\langle sSFR\rangle$,
and we would over-estimate the impact of preprocessing in the 
group environment.

\subsection{MIR Butcher-Oemler Effect}\label{secBO}
\begin{figure}
\epsscale{1.0}
\plotone{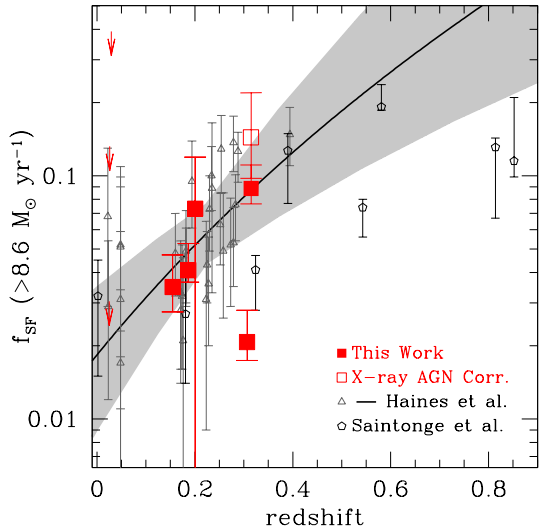}
\caption{Integrated fractions of SFGs ($f_{SF}$, $SFR>8.6 M_{\odot} yr^{-1}$)
in all 8 clusters as a function of redshift.
{\it Filled red squares}
mark the clusters in our sample for which we successfully measure $f_{SF}$,
and {\it red arrows} mark the clusters for which we can produce only
upper limits.  {\it Open red squares} mark the $f_{SF}$ that would be
inferred from X-ray only AGN identifications.  The filled and open 
squares overlap for all clusters except AC114.
{\it Open grey triangles} indicate the LoCuSS clusters as
reported by \citet{hain09}, and {\it open black pentagons} 
mark the clusters measured
by \citet{sain08}.  The solid line indicates the best-fit $f_{SF}$--$z$ 
relation from \citet{hain09}, and the shaded
region shows the $1\sigma$ confidence interval for their fit
($f_{SF}\propto(1+z)^{5.7^{+1.8}_{-1.7}}$). 
\label{figSfrEvol}}
\end{figure}

The relative importance of gas starvation and RPS is also probed by
the evolution in $\langle SFR\rangle$ as a function of cosmic time.  The
classic example of this is the Butcher-Oemler effect
\citep{butc78}.  \citet{hain09}
constructed an analogous measurement with SFRs measured via 
$\nu L_{\nu}(24\mu m)$
among the LoCuSS cluster galaxies.
They employed a SFR threshold of $8.6~M_{\odot}\ yr^{-1}$, and
they found that
$f_{SF}\propto(1+z)^{n}$ with $n=5.7^{+2.1}_{-1.8}$.
Figure \ref{figSfrEvol} shows their fit to $f_{SF}$ among the LoCuSS clusters
as a function of redshift.  The $f_{SF}$ values for our clusters and for
a higher redshift cluster sample measured by
\citet{sain08} are superimposed.
The 8 clusters in our sample, shown as the red triangles in Figure 
\ref{figSfrEvol}, are clearly consistent with the \citet{hain09}
result within the uncertainties.  However, the fit to the LoCuSS clusters
systematically overpredicts $f_{SF}$ in the \citet{sain08}
clusters, despite the lower SFR threshold ($5~M_{\odot}\ yr^{-1}$) used
by \citet{sain08}.

In \S\ref{secSubstr}, we considered the impact of X-ray only AGN
identification on the inferred $\langle sSFR\rangle$.
This becomes a more important consideration at high-$z$,
because the frequency of luminous AGNs increases dramatically \citep{mart09}.
Figure \ref{figSfrEvol} includes two points for each cluster
in our sample.  One shows $f_{SF}$ with the IR AGN selection included
({\it filled triangles}), and the other shows $f_{SF}$ that we would measure
if we only knew about the X-ray selected AGNs ({\it open triangles}).  The 
$f_{SF}$ inferred from the X-ray only selection in AC114 differs by $1.6\sigma$
from the result when the full AGN sample is considered.
This illustrates the contamination that X-ray only AGN identification
can introduce to integrated SFRs.  This contamination becomes 
more severe, and appears in other clusters, for SFR thresholds less than
the fairly high value employed by \citet{hain09}.

\section{Discussion}\label{secDiscuss}
In \S\ref{secPartial} and \S\ref{secSfrRad} we examined correlations 
between environment, SFR and $M_{*}$.  We found a strong correlation of
SFR with $R/R_{200}$, and we found evidence for a transition population
of low-SFR galaxies near the cluster center.  
We interpret this population as evidence 
that galaxies in this region experience a rapid reduction in SFRs
that initiates their transition from SFGs
to passive galaxies.  This interpretation was supported by
a possible trend toward larger $L_{TIR}^{*}$ farther out in the
cluster (\S\ref{secTirLF}).  We also
found evidence for a concentration of massive galaxies
near the cluster center (\S\ref{secMassRad}).

In this section, we consider the results of 
\S\ref{secSfrRad}-\ref{secSubstr} in more detail and interpret them in
the context of two competing mechanisms to end star formation in cluster
galaxies: RPS and gas starvation (\S\ref{secClusterSf}).  
We also briefly discuss the additional information that the 
Butcher-Oemler Effect can provide about the impact of the cluster
environment on SFGs
(\S\ref{secEvolution}).

\subsection{Star Formation in Clusters}\label{secClusterSf}
In \S\ref{secResults}, we examined several diagnostics for the impact of
the cluster environment on star formation.  These include partial
correlation analysis, $\langle SFR\rangle$ and $\langle sSFR\rangle$
versus radius,
and an examination of TIR LFs.  The partial correlation
results informed much of our subsequent analysis.  One important
result was the
absence of a correlation between SFR and $\Sigma_{10}$ once we control
for $R/R_{200}$.  This implies that interactions between individual galaxies
have only a limited
impact on star formation in cluster members.  We conclude that
the SFRs of cluster members are controlled by
hydrodynamic interactions between galaxies and the ICM.

There is disagreement in the literature concerning the importance
of different mechanisms to shut down star formation in clusters.
For example, \citet{sima09} determined that evolution in cluster 
SFRs is controlled by galaxy-galaxy interactions
because the growth in the fractions of early-type and passive galaxies
track one another very closely in their sample.  This contrasts sharply
with our results in \S\ref{secPartial}, which suggest that interactions
with the ICM are the dominant factor.  Of the hydrodynamic processes
commonly considered (e.g.\ RPS and gas starvation), only RPS has been
directly observed to work in nearby clusters \citep{kenn04,siva10}.
We therefore ask whether our observations are consistent
with RPS alone or if an additional mechanism is required to explain
the observations.

\citet{treu03} determined that RPS works effectively for Milky Way-like
galaxies in a cluster with $M_{vir}=8\times10^{14} M_{\odot}$
when $R<0.5 R_{200}$.  Our sample is restricted to
projected $R<0.4 R_{200}$, and their cluster mass is similar to the
typical cluster in our sample ($M_{clust}\approx 5\times 10^{14}~M_{\odot}$),
so RPS should act efficiently on most galaxies in our sample.
Nevertheless, some of the obvious signatures of RPS do not appear:
We found no residual correlation of SFR with $M_{*}$
at fixed $R/R_{200}$, which is contrary to the prediction that RPS should
affect low-mass galaxies more strongly.  
Furthermore, we found in \S\ref{secTirLF} that the TIR LF of cluster
galaxies is significantly better fit by a Schechter function with
variable $L_{TIR}^{*}$ and with the Coma cluster's faint-end slope
than by a redshift-corrected field galaxy LF.
This is inconsistent with the
simplest prediction of RPS, which would suggest that the SFG population in
clusters should be well-described by the field galaxy LF \citep{bai06,bai09}.
To interpret this disagreement, we need to know whether it
results from different faint-end slopes between the cluster population and the
field, different $L^{*}_{TIR}$, or some combination of the two.  This
level of detail is impossible given the completeness limit in the present
sample, so the implications of the poor agreement between the cluster and
field SFG populations are ambiguous.  Finally, the best-fit power law to
the SFR--radius relation is consistent---within very large observational 
uncertainties---with the predictions of \citet{book10}.

However, we also found evidence that RPS contributes significantly
to the decline of SFRs among cluster galaxies.  Figure 
\ref{figSfRad} shows that the dependence of SFR on $R/R_{200}$ is driven
largely by a decline in $f_{SF}$ toward the cluster center, and only a small
fraction of the dependence is driven by decline in the SFRs of individual
SFGs.  This is more consistent with RPS than with gas starvation.  Furthermore,
the residual decline in $\langle SFR\rangle_{SFG}$ with $R/R_{200}$ does
not occur smoothly, but appears to set in rapidly at $R\approx 0.1R_{200}$.
The crossing time for this region ($\sim 100~Myr$) is consistent with
RPS as the origin of the transition, but the small size of this region compared
to the radius where RPS is theoretically expected to be 
important ($0.5~R_{200}$) is surprising.  This explanation is also difficult
to reconcile with the decline in $f_{SF}$, which we attribute to
RPS, across all radii.

To interpret Figures \ref{figSfRad} and \ref{figRadialLF}, 
we must consider projection effects and the influence of galaxy
``backsplash''.
Projection effects will cause some galaxies at intrinsically large $R/R_{200}$
to appear at small radii when the cluster is projected onto the plane of
the sky.  If we assume an $R^{-2}$ density profile
for galaxies and $f_{SF}$ vs.\ $R/R_{200}$ as determined by \citet{vdLi10},
only $\sim 30\%$ of SFGs with projected $R<0.3 R_{200}$ actually fall within
that region.  This represents an upper limit, since the 
\citet{vdLi10} result also includes projection effects.  Therefore, we 
cannot assume that SFGs at small $R/R_{200}$ actually reside in
the high density regions where RPS is most important.  This suggests that
SFGs that physically reside inside $0.1 R_{200}$ have their SFRs reduced
more drastically than implied by Figure \ref{figSfRad}.

``Backsplash'' refers to galaxies on nearly radial orbits that pass 
through the dense
central region of the cluster and return to large $R/R_{200}$.
This effect can make radial gradients like the ones shown in Figure
\ref{figSfRad} particularly difficult to interpret, because even galaxies
presently at large radii may have passed near the cluster center in the past.
\citet{gill05} use N-body simulations to find that
50\% of galaxies between 1-2$R_{200}$ are backsplash galaxies, and 90\%
of these have been inside $0.5R_{200}$ at some point in the past.  
\citet{pimb11} use mixture modeling with observations of real clusters
from SDSS to infer that $60\pm6\%$ of galaxies at $R/R_{200}=0.3$ are
part of the backsplash population, so they were even deeper into the
dense central region of the cluster at some point in the past.  
Galaxies on radial
orbits in a cluster with an $R^{-2}$ mass profile and a core radius
of $0.05 R_{200}$ spend only 4\% of their time inside $0.1 R_{200}$
and only 32\% of their time inside $0.5 R_{200}$.  This model also implies
that approximately half of the galaxies in a cluster with an age of 8 Gyr have
previously passed through the cluster center.
As a result, the dynamic nature of the cluster populations smears the 
dependence of star formation on projected radius
relative to the underlying, three-dimensional trends.  This suggests that
the projected trends shown in Figure \ref{figSfRad} are lower limits
to the true, three-dimensional trends.

The effects of projection are potentially just as important.  We adopt the
simple mass profile described above, which implies that $\sim 40\%$ of 
galaxies with projected $R<0.1 R_{200}$ actually reside outside $0.1 R_{200}$.
However, the trend in $f_{SF}$ versus radius (Figure \ref{figSfRad}) implies
that at least 67\% of SFGs that appear with $R<0.1 R_{200}$ in projection
are likely to have three-dimensional radii outside this limit, and 
20\% of SFGs with projected $R<0.4 R_{200}$ will lie outside this radius.  
The steep
dependence of $f_{SF}$ on radius therefore works to mitigate any radial
trends among the SFG population, which would lead to the small dependence of
$\langle SFR\rangle_{SFG}$ on $R/R_{200}$.  This suggests that the sharp
drop in $\langle SFR\rangle_{SFG}$ at projected radius $0.1 R_{200}$ would
be stronger if measured relative to physical radius.

An ideal way to account for both projection and backsplash in our
observables is to
compare our results to models that include
these effects.  This approach allows more reliable conclusions than
simple, {\it ad hoc} arguments.  \citet{book10} developed
a model for the removal of hot gas from galaxies by the ICM, which is the
physical mechanism that drives gas starvation.  Their ``shocks''
model predicts that galaxies that experience this process
should show $SFR\propto (R/R_{200})^{\sim 0.6}$ between 0.1-0.4$R_{200}$
(their Figure 3).
This is consistent with our results in \S\ref{secSfrRad}
($\langle SFR\rangle \propto (R/R_{200})^{1.3\pm0.7}$).  However, the
model yields overall poor agreement with the data, even after we fit the 
normalization of the model to best match the observations.  This contrasts
with the results of \citet{book10}, who found that their model agrees
well with the SFR--radius relation measured among the CNOC clusters
\citep{balo00}.  The resolution of this conflict will require additional
observations and more sophisticated theoretical models that provide
predictions for competing processes.

The large statistical uncertainties on the measured SFR versus $R/R_{200}$
preclude detailed comparisons between our observations
and a model for either RPS or gas starvation, so we must rely on
qualitative arguments.
Figure \ref{figStackedLF} demonstrates that the field galaxy TIR LF at
the median redshift of the combined cluster galaxy sample provides
a poor match to the observed TIR LF, and this discrepancy is most
pronounced among galaxies with the lowest SFRs.
The bottom panel of Figure \ref{figSfRad} also shows a sudden decrease in the
SFRs of SFGs with projected radii $R<0.1 R_{200}$.  The crossing time for
this region is $\sim 200$~Myr, which suggests that RPS is responsible for
this reduction.  However, only 5\% of SFGs in the cluster have
$R<0.1 R_{200}$, so this reduction in the SFRs of SFGs near the cluster
center cannot account for the discrepancy between the TIR LFs of cluster
and field galaxies.  There must also be an effect at larger radii
that is not apparent in the present sample.  The entire sample has
$R<0.5 R_{200}$, so either RPS or gas starvation could plausibly cause
the disagreement.

In contrast to our results,
\citet{bai09} found that the TIR LFs of many clusters are consistent with
one another and with the field LF.  They inferred that cluster galaxies
only rarely occupy a transition phase
between SFRs characteristic of field galaxies and complete passivity.
From this, they determined that star formation in cluster galaxies
must be truncated on short timescales compared to the
lifetime of the cluster.  Both we and \citet{bai06} find 
$\sim1\sigma$ variations in the shape of the TIR LF in different 
$R/R_{200}$ bins.  The observed decreases in $L_{TIR}^{*}$, while not
statistically significant, are consistent with the decline in
$\langle SFR\rangle_{SFG}$ for $R<0.1R_{200}$
(Figure \ref{figSfRad}).  The latter result indicates
that the high ICM density near cluster centers reduces SFRs in
individual SFGs.  The smooth decline in $f_{SF}$ as a function of
$R/R_{200}$ suggests that this process eventually results in the end of
star formation in these galaxies.
Projection effects and backsplash will both influence the observed trends.
Projection would cause us to mistake galaxies at large $R/R_{200}$ for
galaxies near the cluster center, while backsplash would move galaxies
that had been processed near the cluster center back to the outskirts of
the cluster.  Both effects would cause the projected trends to appear
weaker than the true, three-dimensional variations in the cluster.
This suggests that the observed radial variations are real, and the
trends with projected radius likely underestimate 
the intrinsic, three-dimensional trends.

We find lower $\langle SFR\rangle_{SFG}$ inside $0.1 R_{200}$ compared to
outside this radius.  We also see hints of this change in the lower 
$L_{TIR}^{*}$ near the cluster center compared to further out.
The variation in the properties of SFGs implied by these measurements 
indicates a substantial change in the SFGs very close to the cluster center.
The crossing time for the sphere with radius $0.1 R_{200}$ is less than 200 Myr,
which strongly favors RPS as an explanation.  However, we also see indications
for a deficit in the number of low-SFR cluster galaxies relative to the
field population.
Because only $\sim 5$\% of SFGs have projected $R<0.1 R_{200}$, galaxies 
outside $0.1 R_{200}$ must dominate the
under-abundance of galaxies with $SFR\approx 3 M_{\odot}~yr^{-1}$.

We can determine whether RPS or gas starvation are more likely to be 
responsible for the deficit from the time required to make the transition.
We find that $66$\% of SFGs with
$M_{R}<-20$~mag have $L_{TIR}<L_{TIR}^{thresh}$, where
$L_{TIR}^{thresh}$ is the luminosity expected for a typical field spiral
with $M_{R}=-20$.  If 50\% of field SFGs with the same $M_{R}$ distribution
had $L_{TIR}<L_{TIR}^{thresh}$, then 16\% of cluster SFGs would be
in transition.  Combined with the gas consumption timescale of a typical
spiral galaxy (2.4 Gyr; \citealt{bigi11}), this implies a transition 
time of $\sim400~{\rm Myr}$.  This timescale is approximately twice the 
dynamical time of an ordinary spiral galaxy.  This timescale favors RPS
as the dominant mechanism among the full cluster sample,
because RPS implies that galaxies should remain in a transition phase for 
approximately their dynamical time while the cold ISM is stripped.
However, the assumption that 50\% of field SFGs in an $M_{R}$-matched
sample would have 
$L_{TIR}<L_{TIR}^{thresh}$ is arbitrary.  A comparison
of the SFR--$M_{R}$ relations in clusters and in the field
is required to measure the transition time more precisely.

If star formation in most cluster galaxies ends as a result of RPS,
post-starburst galaxies should be more frequent in clusters than in the
field.  This is a robust prediction of any scenario that results in a rapid
transition of SFGs to passive evolution.  Galaxies with K+A
spectra, which are usually associated with post-starburst populations,
should remain visible for $\sim 100$~Myr to 1 Gyr.  
This is short compared to the
cluster crossing time, so a large population
of K+A galaxies relative to SFGs 
would be strong evidence that RPS plays an important role.
Instead, \citet{yan09} report that galaxies with 
K+A spectra are less common in overdense
environments like clusters than in the field at $z\approx 0.1$.
They suggest that K+A galaxies appear at constant absolute density, and that
this density corresponds to the group scale at $z\approx0$.
\citet{dres99} instead found that the fraction of K+A galaxies is much
higher in clusters than in the field, and \citet{vdLi10} found no
dependence of the ratio of $N_{K+A}/N_{SF}$ on $R/R_{200}$.  The
different methods used by these authors to select their K+A 
samples---\citet{dres99} rely on [O{\sc ii}] to exclude SFGs, while
\citet{yan09} use H$\beta$, and \citet{vdLi10} select galaxies with excess
Balmer line absorption from their principal component analysis---may 
account for
the apparent contradictions in these observational results.

Groups that have recently fallen into a cluster might appear as an excess
in the substructure parameter \citep{dres88}.
In our sample only A3125 has
a mass comparable to galaxy groups, and we have not considered that cluster
in our analysis.  Therefore, we cannot directly constrain the mechanism that
drives SFR evolution in group members.
However, we find that $\langle sSFR\rangle$ is higher among clusters
with no substructure than in A3128, which is the only member of our 
main sample with significant substructure.  This could indicate that
galaxies that have recently been part of groups have
experienced preprocessing, but it could also arise from cosmic 
variance.
In a recent study of SDSS galaxy
clusters, \citet{vdLi10} found a trend of SFR with radius that extended
to $2R_{200}$.  They concluded that preprocessing of galaxies before they
become cluster members is likely to contribute significantly to the 
SFR--radius relation.  Additional observations are required to determine
if galaxy groups are responsible for the preprocessing or if other
processes are required.

\subsection{Evolution}\label{secEvolution}
In \S\ref{secBO} we suggested that the evolution of star formation
in clusters is sensitive to the mechanism(s)
responsible for the appearance of the $z=0$ SFR--density relation.
In particular, the rate of evolution of $f_{SF}$ is sensitive to the
operation of the cluster environment on recently accreted field galaxies.
Figure \ref{figSfrEvol} shows the fraction of SFGs in clusters as a function
of redshift since $z\approx 0.8$ for the clusters in our sample (\textit{red})
compared to the samples of \citet{sain08} and \citet{hain09}.  

We use measurements of $f_{SF}$ versus $z$ to estimate the time required 
for $f_{SF}$ in clusters
to decline by a factor of $e$ compared to coeval field galaxies.
The best-fit to the \citet{hain09} galaxies $\bigl(f_{SF}\propto (1+z)^{m}$,
$m=5.7^{+2.1}_{-1.8}\bigr)$ is shown as the black line in 
Figure \ref{figSfrEvol}, and it
agrees well with the clusters in our sample.  \citet{lefl05}
report that the field galaxy LF evolves as 
$L_{TIR}^{*}(z)\propto(1+z)^{n}$, where $n=3.2^{+0.7}_{-0.2}$.
Approximately 70\% of cluster member galaxies at $z=0$ never had a
massive companion before they entered the cluster environment \citep{berr09},
so we can assume that galaxies that fall into the cluster have the same
LF as field galaxies.  The threshold we use to identify SFGs
($SFR>8.6~M_{\odot}~yr^{-1}$) is larger than the SFR that corresponds to
$L_{TIR}^{*}$ ($4~M_{\odot}~yr^{-1}$), so we assume that $f_{SF}$
among field galaxies has the same redshift dependence as $L_{TIR}^{*}$.
With this assumption, we can measure the relative change in $f_{SF}$ as a
function of redshift and determine how the cluster environment induces
SFGs to turn passive.
The ratio of $f_{SF,clust}$ to $f_{SF,field}$
has undergone approximately $1.7\pm1.2$ $e$-foldings since $z=1$.  
The elapsed time
over this redshift interval is 7.7 Gyr, so the $e$-folding time
for $f_{SF,clust}/f_{SF,field}$ is $4.6^{+10.6}_{-1.8}$~Gyr.

The $e$-folding time of $f_{SF}$ does not correspond
directly to the truncation time for star formation in individual cluster
members.  New SFGs constantly fall into the cluster from the field, and
this results in a longer timescale for $f_{SF,clust}/f_{sf,field}$ to decline
than for SFRs to decline in individual galaxies.  The rate at which SFGs
fall into the cluster combines with the timescale for the conversion
of individual SFGs to passive evolution to determine how rapidly
$f_{SF,clust}/f_{SF,field}$ changes.
The timescale for this change is long compared to the gas exhaustion time
in typical spiral galaxies (2.4~Gyr, \citealt{bigi11}),
but this does not necessarily indicate that
the timescale for the evolution of individual SFGs is similarly long.
If the timescale for evolution of individual SFGs is indeed long, it
would favor gas starvation over RPS as the primary mechanism to end
star formation in cluster galaxies.

In addition to the degeneracy between changes in rates of infall and the
timescale for individual SFGs to stop forming stars, the measured
timescale for $f_{SF,clust}/f_{SF,field}$ to evolve includes significant
observational uncertainty.
The \citet{hain09} best-fit overpredicts $f_{SF}$ among
the high-$z$ clusters, despite the higher SFR threshold employed by
\citet{hain09}, so it underestimates
the timescale over which $f_{SF,clust}/f_{SF,field}$ evolves.
Additional observations are
required to correct this bias.
A measurement of $f_{SF}$ versus $z$ with a longer redshift baseline and
consistent identification of SFGs will appear in our next paper.

\section{Summary and Conclusions}\label{secConclusion}
We have used visible to MIR observations of 8 low-$z$
galaxy clusters to constrain the impact of
the cluster environment on star formation.  
We examined the relationship between star formation and environment
among cluster members and found a strong correlation,
with $\langle SFR\rangle \propto (R/R_{200})^{1.3\pm0.7}$, and this simple
power law provides a good match to the data.
\citet{book10} model the impact of gas starvation on star formation.
Their model is marginally consistent with the power-law that best fits the
observed SFR--radius relation, but it is a poor fit to the data themselves.
The $\langle SFR\rangle$--$R/R_{200}$ relation is dominated
by a decline in the fraction of SFGs toward the cluster center, but we
also find lower $\langle SFR\rangle$ among SFGs with projected $R<0.1 R_{200}$.
The dominance of the decline in $f_{SF}$ and the short crossing time of a 
sphere with radius $0.1 R_{200}$ both suggest that RPS contributes 
significantly to the observed trend in SFR with $R/R_{200}$.
The TIR LFs hint at a shift toward lower $L_{TIR}^{*}$ near the cluster center,
which is consistent with the observed decline in $\langle SFR\rangle$,
but this shift is not statistically significant.
Projections effects and backsplash both work to weaken the observed trends
relative to the intrinsic variation in three dimensions, which can
hide the steep gradients that would be expected from RPS.

We also examined the relationship between $R/R_{200}$ and stellar mass
in cluster galaxies.  We found that galaxies with $R\lesssim 0.1R_{200}$
show larger $\langle M_{*}\rangle$ than galaxies farther out in the cluster,
even after we have eliminated BCGs from our sample.  This excess is significant
at $\sim 3.5\sigma$, and projection effects are also expected to weaken
the observed trend relative to the intrinsic, three-dimensional variation,
so we conclude that it is robust.
Von der Linden et al.\ (2010) found no such excess once they had removed
BCGs, so our result conflicts with theirs.  This difference may result
from the SDSS fiber collisions.  Our sample is limited to galaxies
more luminous than the SDSS $r$-band magnitude limit at the median redshift
of the \citet{vdLi10} cluster sample, so our sample is on average more massive
than theirs.  However, the timescale for dynamical
friction to affect cluster members is much longer than the Hubble time.
This suggests that cluster galaxies undergo mass segregation via virial
relaxation, analogous to the mass segregation exhibited by some
Galactic globular clusters.

We measured the fractions of SFGs in our cluster sample
as a function of redshift,
and we found that these fractions are consistent with the measurements
made by \citet{hain09} for the LoCuSS clusters.  However, incomplete
AGN subtraction can introduce significant contamination to the integrated
star formation in galaxy clusters.  
For example, we found that eliminating only X-ray AGNs from the sample
prior to calculation of $f_{SF}$ results in a $\sim 1\sigma$ excess in AC 114
in Figure \ref{figSfrEvol}.  The consequences are both more significant
and more widespread for lower SFR thresholds.
This can bias measurements of star
formation as a function of redshift, since the AGN contribution is expected
to be more significant at higher redshift \citep{mart09}.
With a long enough redshift baseline,  evolution in $f_{SF}$
with cosmic time can probe the timescale for the
end of star formation in cluster galaxies.  Present observations favor
gas starvation over RPS, but these include important systematic
uncertainties.

On balance, our measurements are most consistent with RPS as the primary
mechanism to reduce star formation in cluster galaxies.
The SFR--radius relation agrees better with the predictions of RPS than with
gas starvation over the range of radii that we study.  This supports the
conclusions of \citet{bai09}, who studied star formation in clusters over
a similar range of radii and determined that RPS dominates the reduction
of SFR among cluster galaxies.  The observation that SFR does not correlate
strongly with $M_{*}$ among SFGs in clusters can be explained by the
concentration of massive galaxies near the cluster center, where RPS
operates most efficiently.
Furthermore, \citet{bai09} also found that the luminous end of 
the TIR LF does not 
vary significantly between $z=0$ clusters, and we similarly find that
the redshift-appropriate field galaxy LF provides a good match to the observed
TIR LF in each cluster.  While this is also consistent with RPS as the primary
mechanism to end star formation among cluster galaxies, the disagreement
between the field galaxy LF and the stacked cluster sample suggests that
there are small deviations between the individual clusters and the field
galaxy LF that we overlook due to limited precision in the measured LFs.
Such deviations would be more consistent with gas starvation than with
RPS.
Additionally, the long timescale we infer for
the evolution of $f_{SF}$ conflicts with the conclusion that RPS ends
star formation among cluster members and is more consistent with gas 
starvation.  This evidence agrees with the conclusions of
\citet{verd08} and \citet{vdLi10}, who found
independent evidence in favor of gas starvation.
A measurement of the rate of change in $f_{SF}$ as a function of redshift
can provide an additional line of evidence to help resolve this disagreement.
Present results favor a long timescale, but these include significant
systematic uncertainties.
A measurement with a single sample of uniformly analyzed clusters will be the 
subject of our next paper.

\acknowledgements
We are grateful to Kim-Vy Tran and Dan Kelson for insightful comments on
an earlier version of this paper and to John Mulchaey for helpful discussions.
DWA thanks The Ohio State University for support via the Dean's Distinguished
University Fellowship.
PM is grateful for support from the NSF via award AST-0705170.
This work is based in part on observations made with the {\it Spitzer Space
Telescope}, which is operated by the Jet Propulsion Laboratory, California
Institute of Technology under a contract with NASA. 
Support for this work
was provided by NASA through an award issued by JPL/Caltech.
This research has made use of the NASA/IPAC Extragalactic Database (NED)
which is operated by the Jet Propulsion Laboratory, California Institute of
Technology, under contract with the National Aeronautics and Space
Administration.

\appendix
\section{Asymmetric Distribution Function}\label{appAsymGaus}
In order to apply the method described in \S\ref{secLF} to construct a
luminosity function, we must smoothly distribute the weight of a galaxy
across the specified luminosity bins.  The method described in 
\S\ref{secBolCorr} to calculate $L_{TIR}$ produces
asymmetric uncertainties, $L_{TIR}=\mu^{+\sigma_{u}}_{-\sigma_{l}}$,
so we need an asymmetric probability density function (PDF) to distribute
weights correctly.  This PDF must reduce
to the Normal distribution in the case when the upper and lower
luminosity uncertainties are equal (i.e.\ Gaussian errors).  
Here, we describe a piecewise smooth function that satisfies these 
requirements.

First, we define an effective dispersion 
$\sigma_{e}=\sqrt{\sigma_{l}\sigma_{u}}$,
where $\sigma_{u}$ and $\sigma_{l}$ are the upper and lower uncertainties
on $L_{TIR}$,
respectively.  We then define an alternative dispersion, $\sigma(L)$,
which describes the instantaneous shape of the PDF at a luminosity $L$,
\begin{equation}\label{eqSigAlt}
\sigma(L) = \begin{cases}
     \sigma_{l} & \text{IF}\ L < \mu-\sigma_{l} \\
     \sigma_{e} + (\sigma_{l}-\sigma_{e})\frac{|L-\mu|}{\sigma_{l}} & \text{IF}\ \mu-\sigma_{l}\leq L<\mu \\
     \sigma_{e} & \text{IF}\ L=\mu \\
     \sigma_{e} + (\sigma_{u}-\sigma_{e})\frac{|L-\mu|}{\sigma_{u}} & \text{IF}\ \mu < L \leq \mu+\sigma_{u} \\
     \sigma_{u} & \text{IF}\ \mu+\sigma_{u} < L
  \end{cases}
\end{equation}
where $\mu$ is the best estimate $L_{TIR}$; $\sigma_{u}$ and $\sigma_{l}$ are
the upper and lower uncertainties on $\mu$, respectively.
$\sigma(L)$ smoothly connects the low-L and high-L tails of
the desired distribution function.
Given $\sigma(L)$, we can calculate the probability density
for a galaxy with measured luminosity $\mu$ at $L$.
This probability density is given by,
\begin{equation}\label{eqAsym}
f(L,\mu,\sigma_{u},\sigma_{l})=\frac{1}{\sqrt{2\pi}~\sigma(L)} e^{-(\mu-L)^{2}/2\sigma^{2}(L)}
\end{equation}
where $\sigma(L)$ is given by Eq.\ \ref{eqSigAlt}.

The PDF described by Eqs.\ \ref{eqSigAlt} and \ref{eqAsym} approaches Gaussian
at the high- and low-L extremes, with dispersions $\sigma_{u}$ and $\sigma_{l}$,
respectively.
It also smoothly connects these two limiting cases, integrates to unity, and
has dispersion equal to the geometric mean of $\sigma_{l}$ and $\sigma_{u}$
at the nominal luminosity.  It therefore gives a PDF for the luminosity
of a given galaxy that satisfies our requirements and that is
consistent with the available information about $L_{TIR}$.

\bibliographystyle{apj}
\bibliography{apj-jour,masterBiblio}

\clearpage
\input{tab1.tex}

\clearpage
\input{tab2.tex}

\clearpage
\input{tab3.tex}

\clearpage
\input{tab4.tex}

\end{document}

%% file: tab1.tex
\begin{deluxetable}{lrrrr}
\tabletypesize{\scriptsize}
\tablewidth{0pc}
\tablecaption{Spectroscopic Completeness
\label{tabComplete}}
\tablehead{\colhead{} & \colhead{$m_{R}$ (Vega)} & \colhead{$R/R_{200}$} &
\colhead{$f_{Cl,spec}$} & \colhead{$\sigma_{f}$}}
\startdata
 a3128 & 14.33 & 0.11 & 0.669 & 0.016 \\ 
       & 15.76 & 0.10 & 0.682 & 0.042 \\ 
       & 16.32 & 0.07 & 1.020 & 0.065 \\ 
       & 16.89 & 0.06 & 0.532 & 0.057 \\ 
 a3125 & 14.60 & 0.17 & 0.000 & 0.000 \\ 
       & 15.92 & 0.22 & 0.000 & 0.000 \\ 
       & 16.29 & 0.05 & 1.000 & 0.000 \\ 
       & 16.77 & 0.16 & 0.000 & 0.000 \\ 
 a2104 & 17.28 & 0.05 & 0.719 & 0.015 \\ 
       & 18.32 & 0.07 & 0.249 & 0.007 \\ 
       & 18.88 & 0.11 & 0.222 & 0.008 \\ 
       & 19.31 & 0.11 & 0.068 & 0.003 \\ 
 a1689 & 17.02 & 0.03 & 0.626 & 0.014 \\ 
       & 18.46 & 0.08 & 0.409 & 0.014 \\ 
       & 19.07 & 0.09 & 0.217 & 0.015 \\ 
       & 19.65 & 0.12 & 0.411 & 0.033 \\ 
ms1008 & 19.10 & 0.08 & 0.693 & 0.054 \\ 
       & 19.79 & 0.04 & 0.293 & 0.013 \\ 
       & 20.35 & 0.08 & 0.225 & 0.014 \\ 
       & 20.91 & 0.11 & 0.084 & 0.010 \\ 
 ac114 & 19.00 & 0.10 & 0.744 & 0.069 \\ 
       & 19.84 & 0.06 & 0.925 & 0.072 \\ 
       & 20.37 & 0.13 & 0.399 & 0.050 \\ 
       & 20.83 & 0.14 & 0.913 & 0.353 \\ 
\enddata
\tablecomments{A sample of spectroscopic completeness measurements as described
in \S\ref{secSpecComp}.
The complete table is available from the electronic edition of the journal.
A brief sample is shown here for guidance regarding form and content.}
\end{deluxetable}

%% file: tab2.tex
\begin{deluxetable}{lrrrrr}
\tabletypesize{\scriptsize}
\tablewidth{0pc}
\tablecaption{MIR Completeness
\label{tabMirComplete}}
\tablehead{\colhead{} & \colhead{$R/R_{200}$} & \colhead{$f_{\nu}(8\mu m)~[Jy]$} &
\colhead{$C_{8\mu m}$} & \colhead{$f_{\nu}(24\mu m)~[Jy]$} &
\colhead{$C_{24\mu m}$} \\
\colhead{(1)} & \colhead{(2)} & \colhead{(3)} & \colhead{(4)} & \colhead{(5)} & 
\colhead{(6)}}
\startdata
 A3128 & 0.06 & $2.65\times 10^{-4}$ & $0.00^{+0.15}_{-0.00}$ & $1.70\times 10^{-4}$ & $0.00^{+0.15}_{-0.00}$ \\ 
       & 0.06 & $8.44\times 10^{-4}$ & $0.51^{+0.18}_{-0.17}$ & $5.41\times 10^{-4}$ & $0.00^{+0.15}_{-0.00}$ \\ 
       & 0.06 & $2.69\times 10^{-3}$ & $0.55^{+0.14}_{-0.20}$ & $1.64\times 10^{-3}$ & $0.31^{+0.16}_{-0.17}$ \\ 
       & 0.06 & $8.57\times 10^{-3}$ & $0.55^{+0.14}_{-0.20}$ & $5.44\times 10^{-3}$ & $1.00^{+0.00}_{-0.19}$ \\ 
       & 0.06 & $2.66\times 10^{-2}$ & $0.55^{+0.14}_{-0.20}$ & $1.74\times 10^{-2}$ & $1.00^{+0.00}_{-0.19}$ \\ 
       & 0.06 & $8.48\times 10^{-2}$ & $0.55^{+0.14}_{-0.20}$ & $5.51\times 10^{-2}$ & $1.00^{+0.00}_{-0.19}$ \\ 
       & 0.06 & $2.73\times 10^{-1}$ & $0.55^{+0.14}_{-0.20}$ & $1.64\times 10^{-1}$ & $1.00^{+0.00}_{-0.19}$ \\ 
       & 0.06 & $8.67\times 10^{-1}$ & $0.55^{+0.14}_{-0.20}$ & $5.49\times 10^{-1}$ & $1.00^{+0.00}_{-0.19}$ \\ 
 A3125 & 0.05 & $2.53\times 10^{-4}$ & $0.00^{+0.50}_{-0.00}$ & $1.58\times 10^{-4}$ & $0.00^{+0.50}_{-0.00}$ \\ 
       & 0.05 & $7.95\times 10^{-4}$ & $1.00^{+0.00}_{-0.50}$ & $5.09\times 10^{-4}$ & $0.00^{+0.50}_{-0.00}$ \\ 
       & 0.05 & $2.40\times 10^{-3}$ & $1.00^{+0.00}_{-0.50}$ & $1.57\times 10^{-3}$ & $0.00^{+0.50}_{-0.00}$ \\ 
       & 0.05 & $7.89\times 10^{-3}$ & $1.00^{+0.00}_{-0.50}$ & $5.03\times 10^{-3}$ & $0.58^{+0.08}_{-0.08}$ \\ 
       & 0.05 & $2.50\times 10^{-2}$ & $1.00^{+0.00}_{-0.50}$ & $1.60\times 10^{-2}$ & $1.00^{+0.00}_{-0.50}$ \\ 
       & 0.05 & $7.84\times 10^{-2}$ & $1.00^{+0.00}_{-0.50}$ & $4.93\times 10^{-2}$ & $1.00^{+0.00}_{-0.50}$ \\ 
       & 0.05 & $2.56\times 10^{-1}$ & $1.00^{+0.00}_{-0.50}$ & $1.57\times 10^{-1}$ & $1.00^{+0.00}_{-0.50}$ \\ 
       & 0.05 & $8.01\times 10^{-1}$ & $1.00^{+0.00}_{-0.50}$ & $4.87\times 10^{-1}$ & $1.00^{+0.00}_{-0.50}$ \\ 
 A2104 & 0.06 & $3.43\times 10^{-5}$ & $0.00^{+0.06}_{-0.00}$ & $2.25\times 10^{-5}$ & $0.00^{+0.06}_{-0.00}$ \\ 
       & 0.15 & $1.13\times 10^{-4}$ & $0.11^{+0.09}_{-0.06}$ & $6.91\times 10^{-5}$ & $0.03^{+0.07}_{-0.03}$ \\ 
       & 0.24 & $3.47\times 10^{-4}$ & $0.51^{+0.17}_{-0.22}$ & $2.27\times 10^{-4}$ & $0.00^{+0.17}_{-0.00}$ \\ 
       & 0.06 & $3.55\times 10^{-3}$ & $1.00^{+0.00}_{-0.03}$ & $2.21\times 10^{-3}$ & $0.68^{+0.09}_{-0.10}$ \\ 
       & 0.15 & $1.10\times 10^{-2}$ & $0.93^{+0.04}_{-0.07}$ & $6.65\times 10^{-3}$ & $0.73^{+0.09}_{-0.09}$ \\ 
       & 0.24 & $3.63\times 10^{-2}$ & $0.70^{+0.11}_{-0.23}$ & $2.16\times 10^{-2}$ & $0.80^{+0.14}_{-0.21}$ \\ 
 A1689 & 0.06 & $2.31\times 10^{-5}$ & $1.00^{+0.00}_{-0.09}$ & $1.42\times 10^{-5}$ & $0.00^{+0.06}_{-0.00}$ \\ 
       & 0.06 & $7.49\times 10^{-5}$ & $1.00^{+0.00}_{-0.09}$ & $4.56\times 10^{-5}$ & $0.00^{+0.06}_{-0.00}$ \\ 
       & 0.06 & $2.33\times 10^{-4}$ & $1.00^{+0.00}_{-0.09}$ & $1.43\times 10^{-4}$ & $0.04^{+0.08}_{-0.03}$ \\ 
       & 0.06 & $7.17\times 10^{-4}$ & $1.00^{+0.00}_{-0.09}$ & $4.45\times 10^{-4}$ & $0.71^{+0.09}_{-0.13}$ \\ 
       & 0.06 & $2.23\times 10^{-3}$ & $1.00^{+0.00}_{-0.09}$ & $1.45\times 10^{-3}$ & $0.81^{+0.09}_{-0.11}$ \\ 
       & 0.06 & $7.10\times 10^{-3}$ & $1.00^{+0.00}_{-0.09}$ & $4.32\times 10^{-3}$ & $0.81^{+0.09}_{-0.11}$ \\ 
       & 0.06 & $2.35\times 10^{-2}$ & $1.00^{+0.00}_{-0.09}$ & $1.43\times 10^{-2}$ & $0.81^{+0.09}_{-0.11}$ \\ 
       & 0.06 & $7.38\times 10^{-2}$ & $1.00^{+0.00}_{-0.09}$ & $4.54\times 10^{-2}$ & $0.81^{+0.09}_{-0.11}$ \\ 
MS1008 & 0.07 & $2.53\times 10^{-6}$ & $0.00^{+0.11}_{-0.00}$ & $0.00\times 10^{0}$ & $-1.00^{+1.00}_{-1.00}$ \\ 
       & 0.07 & $7.55\times 10^{-6}$ & $0.00^{+0.11}_{-0.00}$ & $4.66\times 10^{-6}$ & $0.00^{+0.11}_{-0.00}$ \\ 
       & 0.07 & $2.37\times 10^{-5}$ & $0.10^{+0.17}_{-0.05}$ & $1.43\times 10^{-5}$ & $0.00^{+0.11}_{-0.00}$ \\ 
       & 0.07 & $7.69\times 10^{-5}$ & $1.00^{+0.00}_{-0.09}$ & $4.63\times 10^{-5}$ & $0.00^{+0.11}_{-0.00}$ \\ 
       & 0.07 & $2.41\times 10^{-4}$ & $1.00^{+0.00}_{-0.09}$ & $1.52\times 10^{-4}$ & $0.00^{+0.11}_{-0.00}$ \\ 
       & 0.07 & $7.67\times 10^{-4}$ & $1.00^{+0.00}_{-0.09}$ & $4.54\times 10^{-4}$ & $0.00^{+0.11}_{-0.00}$ \\ 
       & 0.07 & $2.46\times 10^{-3}$ & $1.00^{+0.00}_{-0.09}$ & $1.43\times 10^{-3}$ & $0.00^{+0.11}_{-0.00}$ \\ 
       & 0.07 & $7.48\times 10^{-3}$ & $1.00^{+0.00}_{-0.09}$ & $4.42\times 10^{-3}$ & $0.00^{+0.11}_{-0.00}$ \\ 
       & 0.07 & $2.24\times 10^{-2}$ & $1.00^{+0.00}_{-0.09}$ & $1.48\times 10^{-2}$ & $0.00^{+0.11}_{-0.00}$ \\ 
 AC114 & 0.06 & $2.43\times 10^{-6}$ & $0.00^{+0.07}_{-0.00}$ & $0.00\times 10^{0}$ & $-1.00^{+1.00}_{-1.00}$ \\ 
       & 0.06 & $7.10\times 10^{-6}$ & $0.00^{+0.07}_{-0.00}$ & $4.48\times 10^{-6}$ & $0.00^{+0.07}_{-0.00}$ \\ 
       & 0.06 & $2.16\times 10^{-5}$ & $0.89^{+0.08}_{-0.11}$ & $1.35\times 10^{-5}$ & $0.00^{+0.07}_{-0.00}$ \\ 
       & 0.06 & $7.16\times 10^{-5}$ & $1.00^{+0.00}_{-0.09}$ & $4.33\times 10^{-5}$ & $0.04^{+0.08}_{-0.03}$ \\ 
       & 0.06 & $2.16\times 10^{-4}$ & $1.00^{+0.00}_{-0.09}$ & $1.37\times 10^{-4}$ & $0.56^{+0.11}_{-0.11}$ \\ 
       & 0.06 & $7.05\times 10^{-4}$ & $1.00^{+0.00}_{-0.09}$ & $4.43\times 10^{-4}$ & $0.56^{+0.11}_{-0.11}$ \\ 
       & 0.06 & $2.20\times 10^{-3}$ & $1.00^{+0.00}_{-0.09}$ & $1.40\times 10^{-3}$ & $0.56^{+0.11}_{-0.11}$ \\ 
       & 0.06 & $6.79\times 10^{-3}$ & $1.00^{+0.00}_{-0.09}$ & $4.28\times 10^{-3}$ & $0.56^{+0.11}_{-0.11}$ \\ 
       & 0.06 & $2.14\times 10^{-2}$ & $1.00^{+0.00}_{-0.09}$ & $1.36\times 10^{-2}$ & $0.56^{+0.11}_{-0.11}$ \\ 
\enddata
\tablecomments{Column (1) gives the cluster name.  Column (2) gives the median 
radius, scaled to the virial radius 
of the cluster, of galaxies that go into the bin.  Columns (3) and (5) give the 
median observed frame fluxes in the $8\mu m$ and $24\mu m$ channels, 
respectively, of the model SEDs that make up each bin.  Fluxes are calculated by
integrating model SEDs with random combinations of the \citet{asse10} star 
forming templates across the published instrument response functions.  
If the SFR inferred from the rest frame luminosities in the model SEDs are
outside the range $10^{-2}<SFR/(1~M_{\odot}\ yr^{-1})<10^{2}$, the associated
fluxes are not included in the sample.  Because the \citet{asse10} templates
are not constructed to have identical SFRs in the $8\mu m$ and $24\mu m$
channels, this sometimes means that an SED with a valid SFR in one channel will
not appear in another.  
When a flux bin is occupied in one channel and not in
another, the empty channel has $f_{\nu}=0$ and $C_{\lambda}=-1$.  This is
the case for the first flux bin in AC114.  Columns (4) and (6) give the
MIR completeness ($C_{\lambda}$) as defined in \S\ref{secMirComp}.
The complete table is available from the electronic edition of the journal.
A sample is shown here for guidance regarding form and content.}
\end{deluxetable}

%% file: tab3.tex
\begin{deluxetable}{ccccccc}
\tabletypesize{\scriptsize}
\tablewidth{0pc}
\tablecaption{Cluster Member Summary
\label{tabGalaxies}}
\tablehead{\colhead{Name} & \colhead{RA} & \colhead{Dec} & \colhead{z} &
\colhead{$M_{*}~[10^{10}M_{\odot}]$} & \colhead{SFR $[M_{\odot}\ yr^{-1}]$} &
\colhead{$\delta$} \\
\colhead{(1)} & \colhead{(2)} & \colhead{(3)} & \colhead{(4)} & \colhead{(5)} &
\colhead{(6)} & \colhead{(7)}}
\startdata
a3128-001 & 03:30:37.7 & -52:32:57.7 & 0.063 & $  3.2\pm  0.7\pm  1.9$ & $< 0.33$ & 0.89 \\
a3125-001 & 03:27:20.2 & -53:28:34.6 & 0.062 & $  3.3\pm  0.7\pm  2.0$ & $  5.1\pm  0.7$ & 3.04 \\
a644-005 & 08:17:25.8 & -07:33:42.5 & 0.071 & $  0.1\pm  0.0\pm  0.1$ & --- & 1.07 \\
a2104-001 & 15:40:07.6 & -03:17:06.8 & 0.153 & $  3.1\pm  1.0\pm  1.9$ & $< 0.20$ & 1.33 \\
a1689-004 & 13:11:29.5 & -01:20:27.7 & 0.183 & $ 67.1\pm 18.2\pm 22.9$ & $< 1.22$ & 1.47 \\
a2163-001 & 16:15:25.8 & -06:09:26.4 & 0.200 & $<  1.4$ & $< 0.50$ & 1.99 \\
ms1008-001 & 10:10:34.1 & -12:39:52.7 & 0.308 & $  5.3\pm  1.7\pm  1.6$ & $< 0.87$ & 0.81 \\
ac114-001 & 22:58:52.3 & -34:46:47.9 & 0.303 & $<  0.1$ & $< 0.39$ & 1.20 \\
\enddata
\tablecomments{The properties of cluster member
galaxies, determined using the methods described by \citet{atle11}.
(1) The name
of this object, which is identical to the name given in Table 2 of
\citet{atle11}.
  (2-3) Positions of each object in J2000 coordinates, as
determined from the $R$-band images of identified cluster members.
(4) Redshifts of each object, as determined by Martini et al.\ (2006,2007)
where available, or from the literature otherwise.  (5) Stellar masses
derived using mass-to-light ratios appropriate for each galaxy's color
and assuming a scaled Salpeter IMF with Bruzual \& Charlot population
synthesis model (\citealt{bell01}, Table 4).  
The first uncertainty quoted
gives the statistical error, and the second gives the systematic error.
Systematic errors are
derived by applying the M/L coefficients appropriate for a Salpeter IMF and
the {\sc{P\'egase}} population synthesis model.  (6) Star-formation rates
either from the $8\mu m$ luminosity, the $24\mu m$
luminosity or by taking the geometric mean of the two, depending on
the measurements available.  Uncertainties include only statistical
errors, and upper limits are quoted at $3\sigma$ in the more sensitive
of the $8\mu m$ and $24\mu m$ bands.  (7) Substructure parameter of
\citet{dres88},
$\delta=(11/\sigma^{2})\times\bigl[(\bar{v}_{local}-\bar{v})^{2}+(\sigma_{local}-\sigma)^{2}\bigr]$,
where the local average velocities ($\bar{v}$) and velocity dispersions
($\sigma$) are calculated over the 10 nearest neighbors of each galaxy.
The complete table is available from the electronic edition of the journal.
A brief sample is shown here for guidance regarding form and content.}
\end{deluxetable}

%% file: tab4.tex
\begin{deluxetable}{cccccccc}
\tabletypesize{\scriptsize}
\tablewidth{0pc}
\tablecaption{Partial Correlation Results
\label{tabSfCorr}}
\tablehead{\colhead{} & \colhead{} & \colhead{Partial $r_{s}$} & 
\colhead{Prob.} \\
\colhead{(1)} & \colhead{(2)} & \colhead{(3)} & \colhead{(4)}}
\startdata
SFR     &   $M_{*}$     & $+0.091$  & $5.15\times 10^{-01}$\\
SFR     &   $R/R_{200}$ & $+0.342$  & $2.11\times 10^{-05}$\\
SFR     &   $\delta$    & $-0.101$  & $5.26\times 10^{-01}$\\
SFR     &   $\Sigma$    & $+0.018$  & $8.40\times 10^{-01}$\\
$M_{*}$ &   $R/R_{200}$ & $-0.087$  & $5.14\times 10^{-01}$\\
$M_{*}$ &   $\delta$    & $-0.024$  & $7.93\times 10^{-01}$\\
$M_{*}$ &   $\Sigma$    & $-0.012$  & $8.90\times 10^{-01}$\\
$R/R_{200}$ & $\delta$  & $+0.188$  & $2.77\times 10^{-02}$\\
$R/R_{200}$ & $\Sigma$  & $-0.576$  & $1.96\times 10^{-17}$\\
$\delta$ &  $\Sigma$    & $+0.068$  & $5.41\times 10^{-01}$\\
\enddata
\tablecomments{Partial correlation results for star forming galaxies
derived from the Spearman
correlation coefficients for the variables listed in columns
(1) and (2).  Column (3) gives the strength of the correlation between
the two variables with the other parameters held fixed.  Column (4) gives the 
probability that a correlation at least as strong as that observed 
might occur by chance among intrinsically uncorrelated data.}
\end{deluxetable}